\documentclass[conference]{IEEEtran}
\IEEEoverridecommandlockouts
\usepackage{cite}
\usepackage{amsmath,amssymb,amsfonts}
\usepackage{algorithmic}
\usepackage{graphicx}
\usepackage{textcomp}
\usepackage{xcolor}
\usepackage{multirow}
\usepackage{float}
\usepackage{subfig}
\usepackage{array}
\usepackage{colortbl}
\usepackage{subfloat}

\def\BibTeX{{\rm B\kern-.05em{\sc i\kern-.025em b}\kern-.08em
    T\kern-.1667em\lower.7ex\hbox{E}\kern-.125emX}}

\begin{document}

\title{On Software Ageing Indicators in OpenStack}

\author{
	\IEEEauthorblockN
	{
		Yevhen Yazvinskyi\textsuperscript{1}, 
		Jasmin Bogatinovski \textsuperscript{1}, 
		Jorge Cardoso\textsuperscript{2,3}, 
		Odej Kao\textsuperscript{1}
	}
	
	\IEEEauthorblockA{
		Technical University Berlin\textsuperscript{1}, Berlin, Germany, email: \{jasmin.bogatinovski, odej.kao\}@tu-berlin.de}
	\IEEEauthorblockA{
		\textsuperscript{2}\textit{Huawei Munich Research}, Munich, Germany }	
	\IEEEauthorblockA{\textsuperscript{3}Departamento de Engenharia Informatica, University of Coimbra, Portugal, email: jorge.cardoso@huawei.com}
		
}

\maketitle

\begin{abstract}
Distributed systems in general and cloud systems in particular, are susceptible to failures that can lead to substantial economic and data losses, security breaches, and even potential threats to human safety. Software ageing is an example of one such vulnerability. It emerges due to routine re-usage of computational systems units which induce fatigue within the components, resulting in an increased failure rate and potential system breakdown. Due to its stochastic nature, ageing cannot be directly measured, instead ageing indicators as proxies are used. While there are dozens of studies on different ageing indicators, their comprehensive comparison in different settings remains underexplored. In this paper, we compare two ageing indicators in OpenStack as a use case. Specifically, our evaluation compares memory usage (including swap memory) and request response time, as readily available indicators. By executing multiple OpenStack deployments with varying configurations, we conduct a series of experiments and analyze the ageing indicators. Comparative analysis through statistical tests provides valuable insights into the strengths and weaknesses of the utilised ageing indicators. Finally, through an in-depth analysis of other OpenStack failures, we identify underlying failure patterns and their impact on the studied ageing indicators.
\end{abstract}


\section{Introduction}

Over the last two decades, the approach to the allocation and distribution of computing capabilities has shifted many fields towards the cloud. Cloud computing is a concept in which computing resources are virtualized and provided as a service~\cite{cloudcomputing_2020}. The inevitable weaknesses in hardware and software make clouds prone to failures, which can lead to economic losses, data losses, security issues, or even endanger human safety~\cite{cloud_risks_2009, cloud_risks_2014}. Because of this, understanding potential performance degradation and the possibilities of potential software failures is important for optimal cloud utilization.

While the cloud can fail in many different ways, in this paper we focus on one specific failure type -- software ageing~\cite{dependabilitydefinition2004}, as a relatively not well-studied area in reliability engineering~\cite{Notoro2021}. Software ageing is a phenomenon in which prolonged usage of complex computational systems leads to the fatigue of their components~\cite{soft_aging_fundamentals_2008}. It can lead to an increased failure rate, degrade system performance, and even result in premature system failures. Software ageing negatively impacts the end user by reducing system performance, it's reliability and forcing administrators to perform maintenance, collaterally reducing system availability as well. Notably, due to the rather vague definition of software ageing, we can not measure the software ageing level itself, but rather we measure its effect on different proxies, referred to as ageing indicators. To mitigate it, software rejuvenation, as a maintenance technique is used~\cite{software_rejuvenation_1995}. 

In the context of cloud systems, memory utilization and response time are the two most commonly used indicators, that are easy to obtain, therefore we focus our study on the two~\cite{aging_rejuv_survey_2020}. Memory consumption is used to calculate time before resource exhaustion, while response time is used as a direct performance degradation metric. Studying the degradation trends in response time and memory consumption against failure frequency can serve to predict an increase in failures. 

As software rejuvenation strategies depend on the selected ageing indicator, related studies often focus on a single indicator analysis~\cite{aging_rejuv_where_we_are_2011}, in single-node cloud deployments. Furthermore, in the context of the cloud, ageing is studied mostly through resource consumption~\cite{aging_eucalyptus_2011_memory, aging_response_hosted_app_2021, response_time_app_aws_2010}. At the same time, Pflanzner et al. showed the importance of OpenStack response time evaluation \cite{openstack_performance_rally_2016}. In their study, they observe OpenStack failures as well as OpenStack response time degradation but do not study it in terms of software ageing. Finally, the failures themselves can affect the software ageing indicators and further interrupt the process (e.g., maintenance activity is conducted before the system fails due to ageing). Therefore, the analysis of the software ageing indicators concerning the failure rate is important, but not well studied, particularly in cases of high concurrency, and multi-node cloud deployments.

Following this, in this paper, we conduct a comprehensive comparison of memory utilization and response time as key ageing indicators for software ageing evaluation. By utilizing response time alongside memory utilization (including swap memory), we aim to explore the effects of software ageing and its implications on system failure rate and thus on system performance and reliability. To conduct our study we selected OpenStack as an open-source solution for deploying private clouds. To conduct the study we designed accelerated experiments on software ageing in single and multi-node deplyoments~\cite{aging_rejuv_where_we_are_2011, aging_rejuv_survey_2020}. We define and deploy an OpenStack configuration, workloads and failure injector. Through the statistical analysis of the measurements, we evaluate how significantly software ageing reduces system response time and memory utilization, how it can potentially worsen the user experience, and how it is connected with rejuvenation.

The contributions of this paper are the following:
\begin{itemize}
    \item \textbf{Failure analysis:} We demonstrate different ways in which OpenStack is affected by failures, how errors play a part in the software ageing process and how rejuvenation influences failures in OpenStack.
    \item \textbf{Ageing evaluation:} We demonstrate how to apply software ageing evaluation on different ageing indicators and analyze their correlation with failures.
    \item \textbf{Ageing indicators evaluation:} We evaluate how software ageing influences response time and memory utilization, the effectiveness of response time as an ageing indicator, and the effectiveness of our ageing evaluation methodology in general.
\end{itemize}

The rest of the paper is organized as follows. In Section~\ref{background} we give an overview of the terminology and related work needed to understand our approach. In Section~\ref{methodology} we describe our methodology including the system we designed, the accelerated experimental setup to simulate software ageing and the details of the analysis. Section~\ref{evaluation} gives extensive details of 12 experimental scenarios we conducted alongside a high-level summary of our findings. Finally, Section~\ref{conclusion} concludes the work. Section~\ref{appendix} gives details on the used statistical tools and the tesetbed we develop to conduct the experiments. 

\section{Background}~\label{background}
In the following, we give a broad background on Openstack and the specific choices of configuration, and software ageing. 

\subsection{Target Cloud System: Openstack}
OpenStack was originally developed by NASA and Rackspace, but over time it has been restructured into an open-source ecosystem consisting of more than 30 different components \cite{openstack_nasa_popular_aws_2013}. Initially, OpenStack was developed as an Infrastructure as a Service software (IaaS), but as many new modules were developed, OpenStack started to be also associated as a Platform as a Service (PaaS) solution.

As a modular system, OpenStack consists of different parts, which are responsible for different tasks. It includes modules for automated cloud deployment, modules for testing system performance by means of artificial load provisioning, and modules for system monitoring. Due to the diverse array of components and the straightforward configurability inherent in OpenStack, the platform offers a comprehensive toolkit conducive to the execution of experimental investigations into ageing phenomena. The availability of open-source code, coupled with the widespread adoption of OpenStack, further enhances its suitability for generating results with real-world applicability. Therefore, we choose OpenStack as our main experimental platform. As a highly complex and modular system, one can derive many different combinations of the core components based on different needs. In this paper, to make the study feasible, we focus on the fundamental OpenStack components that are responsible for its core functionality: compute (Nova), network (Neutron), identity (Keystone), image (Glance), and block storage (Cinder) \cite{openstack_architecture_2014}.

The compute service, also called Nova, provides a way to provision compute instances to OpenStack users by supporting the creation of virtual machines and bare-metal servers, and has also limited support for system containers. Provisioning virtual machines to the user requires images, which have to be provided by an Image Service. A network is also required for the system to be able to launch a compute virtual machine.

There are a lot more different modules and components in the OpenStack ecosystem, many of which require complex configuration processes to ensure proper communication channels between each other. To simplify the configuration and installation process of OpenStack modules, a project called Kolla-Ansible is developed. Ansible is an open-source automation tool for simplifying IT infrastructure management through configuration-driven automation. It is widely used for OpenStack testbed purposes \cite{kolla-ansible_testbed_2021, kolla-ansible_testbed_2020, kolla_rally_2017}. To provide production-ready and deployment tools for operating OpenStack clouds, Kolla-Ansible unites two modern approaches: Ansible orchestration of Docker containers.

OpenStack ecosystem includes a project for performance testing purposes, which is called Rally OpenStack \cite{kolla_rally_2017, openstack_performance_rally_2016}. It is a comprehensive benchmarking framework specifically tailored for assessing and fine-tuning the performance of OpenStack-based cloud environments. To achieve it, Rally provides an easy way to define parameterized scenarios of OpenStack workloads in JSON or YAML format, execute them automatically, and collect workload statistics for the response processing time of different OpenStack modules.

\subsection{Software Ageing in OpenStack}
Software ageing is a phenomenon, that manifests itself as an accumulated degradation of software performance over time as a consequence of various factors such as resource depletion, design flaws, or environmental changes~\cite{software_aging_1994, soft_aging_fundamentals_2008}. Different software fault activations cause software ageing effects by gradually leading the system’s state toward an increased rate of failure occurrences. This gradual trend as a consequence of ageing effects accumulation inside the system is the fundamental difference between software ageing and the system being overloaded.

Due to its nature, software ageing can not be measured or calculated directly. Instead, while the system is in use, ageing effects can be detected, by observing ageing indicators \cite{soft_aging_fundamentals_2008, software_rejuv_hybrid_2013}. Ageing indicators are explanatory variables, which analysis can suggest if the system's performance has degraded or not, and to which level. 

To decrease or prevent the effects of software ageing, predictive maintenance may be used. In the case of software ageing, such a technique is called software rejuvenation \cite{software_rejuvenation_1995}. Software rejuvenation aims to reduce or prevent the impact of software ageing by carrying out various maintenance actions. The scheduling of software rejuvenation can be time-based, inspection-based, or a combination of both. Most software rejuvenation techniques involve restarting or migrating resources. Due to the multilayered nature of OpenStack, software ageing can occur at different levels: ageing can occur inside the operating system, Docker service that runs containers with OpenStack \cite{docker_aging_2019, docker_aging_2021, docker_aging_memory_2020}, or inside OpenStack modules \cite{docker_aging_2021,software_aging_app_in_cloud_2021, docker_aging_2019}. Software rejuvenation is related to different costs at which it is carried out~\cite{rejuv_redundancy_2012}. To properly assess the risks of delayed rejuvenation and, therefore, its cost efficiency, it is necessary to evaluate the effect that software ageing has on the system with the help of the aforementioned software ageing indicators. 

\subsection{SWARE Methodology}
De Melo et al.~\cite{sware_2017_approach_rejuvenation} propose an approach for software ageing and software rejuvenation experiments. This methodology divides the experiment into 3 phases: stress, wait, and rejuvenate. Respectively, during the stress phase the target system is under stress load, during the wait phase it is idle and during the rejuvenation phase, a rejuvenation action is executed, followed by monitoring the system over a certain period. During all 3 phases ageing indicators are observed and analyzed. To analyze the ageing picture in the best possible way, authors deconstructed the difference between overload and ageing manifestation on ageing indicators, as well as the role of rejuvenation in this process. We take this analysis into account during the evaluation of our study, but we do not use the waiting phase, because response time as an ageing indicator can not be measured without the system being used. That said, we see potential in extending our methodology using the modified SWARE approach, in the form of adding another stress phase after the waiting phase.

\section{Methodology}~\label{methodology}
This section gives an overview of our methodology. We build our methodology based on SWARE, and further extend it for our design scenario to enable the monitoring and analysis of memory usage alongside the response time. We further describe the accelerated testing as a precondition to set the system into an ageing state, as well as the analysis details. 

\subsection{OpenStack Accelerated Testing}
To evaluate request response time and memory utilization as indicators of cloud ageing, it is necessary to observe them within an ageing system. To facilitate the rapid and clear demonstration of ageing trends, we aim to expedite the ageing process through the implementation of accelerated testing. The underlying mechanisms of ageing can be attributed to the activation of software faults that progressively impact system performance. These faults can be activated either directly by user-initiated processes or by background operations beyond the user's immediate control, such as automated system cleanups or backups. For the purpose of this study, we focus on the former type of activators, as we can regulate the frequency of user-initiated actions. Through this approach, we aim to conduct accelerated software ageing testing by generating OpenStack user requests with increased intensity.

To serve as the primary ageing actuator, we designate an OpenStack workload, represented by a sequence of diverse requests to the cloud. The central action within our experiment entails the initiation and termination of a virtual machine. The instantiation of a virtual machine necessitates the utilization of a \texttt{flavor}, an \texttt{image}, and an OpenStack network. Here, a flavor denotes the fundamental configuration of the virtual machine, while an image refers to the operating system source intended for deployment within the target virtual machine. These elements are managed by the following OpenStack modules: Nova, Glance, and Neutron. Requests to these modules form the fundament of our workload. 

Thus, our workload involves the execution of requests targeting the following modules: Nova, Glance, Neutron, Cinder, and Keystone. It encompasses a systematic series of requests to OpenStack services, involving the creation, operation, and eventual deletion of diverse entities. In the event of an exception occurring during a workload execution, we opt not to continue the process, as subsequent steps depend on the preceding ones. However, we do not halt the entire workload, as doing so would prevent the removal of OpenStack entities that have already been created. Instead, the workload skips any further steps involved in the creation or operation of new entities and proceeds directly to delete the entities generated before the failed step.

\subsection{Experiment Scenarios Definition}
Another important aspect is to define the experiment scenarios. To that end, we build upon the aforementioned SWARE methodology. We introduce several modifications that enable us to combine the sequential execution of workloads with rejuvenation. The details are given in the following.

The request response time is only measurable during the execution of requests in OpenStack, which itself can act as a potential ageing actuator, while the memory can be obtained during the whole time. Consequently, for the response time, the waiting phase of the SWARE methodology in the traditional sense is not applicable, as we cannot monitor how ageing indicators change during idle system periods. Furthermore, the request-response time cannot be observed during the rejuvenation phase.

To compensate for the lack of data on ageing indicators post the stress phase, we introduce a supplementary post-rejuvenation phase. This phase aims to provide new information for analysis following the completion of the rejuvenation process. Consequently, we define our experimental scenario as comprising three phases: stress, rejuvenation, and post-rejuvenation. In the stress phase, we repeatedly execute workloads on a freshly deployed OpenStack, anticipating an accelerated ageing effect over time due to the continuous execution of ageing actuators. This effect is expected to manifest as a degradation trend in the ageing indicators. To initiate the rejuvenation phase directly after the stress phase, we perform time-based rejuvenation, specifically, a redeployment of OpenStack. While this approach allows us to eliminate ageing accumulated within the cloud, it only partially affects ageing in Docker or the operating system.

The post-rejuvenation phase serves to reveal the state of ageing indicators after the rejuvenation process. It aims to demonstrate the extent of any ageing reversal following rejuvenation. If ageing was present before rejuvenation, we expect OpenStack to exhibit less ageing afterwards. Consequently, the ageing indicators display a partial return of readings, counter to the trend captured during the stress phase. To achieve this, our post-rejuvenation phase involves a shortened stress phase, specifically a brief execution of workloads.

\begin{figure}[!t]
\includegraphics[width=0.5\textwidth]{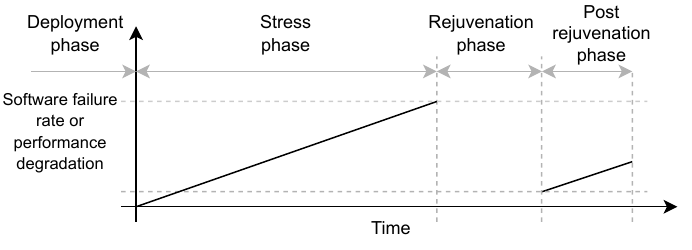}
\caption{Ageing in different phases.}
\label{aging_picture}
\end{figure}

Figure~\ref{aging_picture} illustrates the projected behaviour of the anticipated system state throughout the three phases. It is essential to note that the performance of OpenStack is expected to marginally decline after rejuvenation due to the potential effects of software ageing on Docker and the host operating system. To comprehensively observe the effects of software ageing under various conditions, we have structured the experimental phase into distinct scenarios, each characterized by two key variables: the OpenStack configuration and the level of workload concurrency. Utilizing different levels of concurrency enable us to examine the relationship between workload intensity and ageing. OpenStack is limited by the amount of entities, that exist simultaneously. By raising concurrency above the OpenStack limit, we put the system into an overloaded state, which provide additional insights into the ageing process.

To encompass a broad spectrum of OpenStack load intensities, we employ the following concurrency settings in this study: 1, 2, 4, 8, 16, and 64. Given that the default limit setting in OpenStack is 10, scenarios featuring workload concurrency levels of 16 and 64  correspond to instances of an overloaded system.

This study includes both the all-in-one and multi-node configurations of OpenStack. While a considerable portion of OpenStack ageing studies focuses solely on the all-in-one configuration, the multi-node configuration stands as the primary production setup for OpenStack \cite{guedes2019availability}. Thus, investigating the relationships between different ageing indicators for both configurations offers valuable insights from both academic and practical perspectives on software ageing. It should be noted that we use control node values when analyzing resource consumption. It plays the central role in the entire workload, participating in all crucial actions, thus making it a main target of potential ageing actuators throughout the entire workload. Consequently, we have defined 12 scenarios, combining two types of OpenStack deployments with six levels of workload concurrency. Table \ref{table:scenario_definitions} summarizes the scenarios.

\begin{table}
\begin{center}
\resizebox{0.47\textwidth}{!}{%
\begin{tabular}{ |c|c|c|c|c|c|c|c|c|c|c|c|c| }
  \hline
  Scenario number& 1 & 2 & 3 & 4 & 5 & 6 & 7 & 8 & 9 & 10 & 11 & 12 \\ 
  \hline
  Configuration & \multicolumn{6}{|c|}{multi-node} & \multicolumn{6}{|c|}{all-in-one} \\
  \hline
  Concurrency& 1 & 2 & 4 & 8 & 16 & 64 & 1 & 2 & 4 & 8 & 16 & 64  \\ 
  \hline
\end{tabular}
}
\end{center}
\caption{List of experimental scenarios}
\label{table:scenario_definitions}

\end{table}

\begin{figure*}[!t]
\centering
\includegraphics[width=0.75\textwidth]{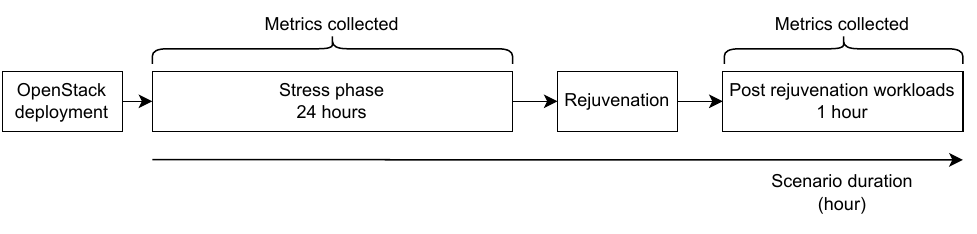}
\caption{Scenario structure.}
\label{figures/chapter_4/scenario_scheme}
\end{figure*}

Experimental studies on software ageing often encompass scenarios of varying durations, ranging from 10 hours to 30 days \cite{openstack_aging_10_hours, araujo2011software}. Considering the diverse configurations of OpenStack and the varying levels of workload concurrency, both of which significantly impact the performance of OpenStack and consequently the intensity of the ageing process, we define the stress phase as continuous workload execution for 24 hours and the post-rejuvenation phase as continuous workload execution for 1 hour. Figure~\ref{figures/chapter_4/scenario_scheme} depicts the course of a single experimental scenario.

\subsection{OpenStack Ageing Analysis}
After the completion of each scenario, various metrics describing the state of OpenStack are obtained. In our ageing analysis, we focus on the following ones: workload duration, memory and swap utilization.

Workload duration represents the total time taken by OpenStack to execute all tasks requested during a single workload. In this way, this metric represents the generalized speed at which the cloud executes user requests. However, it is important to note that not all OpenStack actions included in the workload sequence may be executed due to potential interruptions caused by software failures.

For memory usage, it must be noted, that a more intricate approach is necessary, considering its multifaceted nature. Memory available is the amount of memory which is available for allocation to existing or new processes. The observed reduction in the \texttt{memory available} indicator throughout the OpenStack usage provides a clear trajectory for understanding trends in the overall memory utilization of the system over the experiment's duration. As the system depletes available memory resources, it initiates the use of swap, a disk space acting as an extension of the physical memory. In this way, the \texttt{memory available} metric directly influences \texttt{swap used}, with decreasing available memory leading to heightened swap usage. In instances of continuous performance degradation, we anticipate observing a downward trend in the \texttt{memory available} metric and an upward trend in \texttt{swap used}. However, accessing data from disk is notably slower compared to memory access. Therefore, the combination of these two metrics not only illustrates patterns in memory utilization changes attributed to ageing but also reveals the onset and magnitude of the transition towards employing slower swap memory. This insight allows for the interpretation of a RAM deficit, directly influencing system performance.

OpenStack's failures can manifest in various forms, but our primary focus lies on those that make an exception. Exception errors are particularly valuable as they offer detailed insights into the root causes of failures. Moreover, the exception mechanism grants us the ability to regulate the workflow following the occurrence of an error. When a failure occurs during the execution of a workload, we refer to it as a \texttt{failed workload}, while the specific step at which the error occurs is termed a \texttt{failed step}.

If a failed workload successfully deletes all entities created during its execution, we classify it as a \texttt{non-ageing workload}, and the error causing the interruption is termed a \texttt{non-ageing error}. While every failure has the potential to contribute to ageing, non-ageing errors result in a clean system state, and they have no direct impact on OpenStack ageing in terms of its entities. However, if a failed workload leaves some entities intact after execution, we classify it as an \texttt{ageing workload}, and the error leading to its interruption is referred to as an \texttt{ageing error}. The entities left behind, termed \texttt{leftover entities}, potentially contribute to cumulative resource consumption and have an impact on OpenStack's ability to execute workloads.

Failures can lead to ageing through various mechanisms, for example, entities created with a failed state or connection errors. In instances where a workload step fails but generates an OpenStack entity with an error status, this additional entity remains in the system as our cleanup mechanism is designed to delete entities created before the failed step, not during it. Similarly, connection issues result in the workload's inability to clean leftover entities.

\begin{figure*}[!t]
\centering
\includegraphics[width=0.65\textwidth]{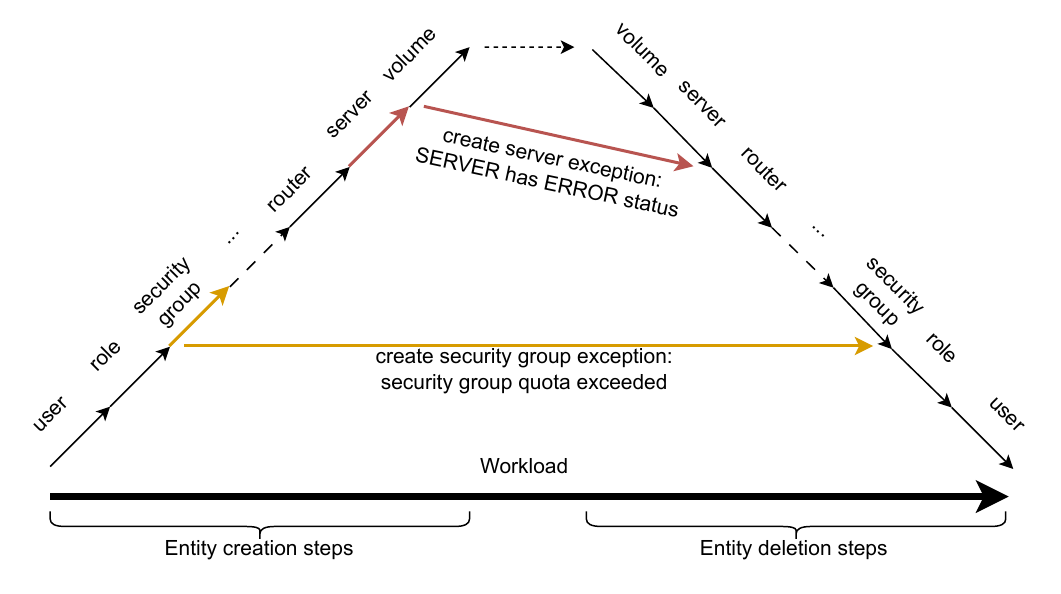}
\caption{Comparison of non-ageing and ageing errors during a workload.}
\label{figure/chapter_6/failures/error_type_comparison}
\end{figure*}

Figure~\ref{figure/chapter_6/failures/error_type_comparison} provides an example of what these two types of errors can look like. \texttt{Security group quota exceeded} exception can occur during the \texttt{Create security group} step, falling under the non-ageing error type, as it does not generate a new OpenStack entity. It leads to the removal of all entities created before the exception, such as roles and users. On the other hand the \texttt{SERVER has ERROR status} exception during \texttt{Create server} step results in the creation of a server entity, followed by the deletion of all entities created before this exception, excluding the server.

Ageing errors directly affect OpenStack's resource ageing as leftover entities, although not in use, continue to consume system resources. Additionally, these errors impact OpenStack's ability to simultaneously operate multiple entities, given the platform's defined limitations, referred to as quotas. In this study, we utilize OpenStack with a default quota configuration, allowing a maximum of 10 Instances, 10 Security groups, 10 Routers, and 10 Volumes to be simultaneously used. For instance, if OpenStack already has 10 existing Security groups and receives a request for an additional one, it would reject the request, producing a \texttt{Security group quota exceeded} error. We define the capacity of OpenStack as the number of workloads it can successfully execute simultaneously in the absence of errors. On a freshly installed OpenStack, the capacity value is determined by the lowest OpenStack entity quota utilized in our workload, in addition to constraints on natural resources, such as available memory or network capabilities.

Each time an ageing error occurs, OpenStack's capacity decreases by one, provided the error leads to the creation of an additional leftover entity of a type that already limits OpenStack's capacity. For example, if OpenStack has a default quota of 10 for all entities and has already accumulated three leftover server instances, its capacity decreases to 7. Consequently, any concurrent workload request exceeding this capacity would trigger an \texttt{Instance quota exceeded} error. If another ageing error occurs, generating a new error router, the capacity of OpenStack would remain unchanged, as the highest number of concurrent workload executions would still be limited to 7 by the instance quota. However, if an additional three ageing errors create another three leftover routers, resulting in four leftover routers in total, the router quota would become the new limitation, reducing OpenStack's capacity to 6.

When OpenStack receives more requests, than it can handle at the same time, we say that it operates in an \texttt{overloaded state} or simply that it is overloaded. Overloading can happen both due to hardware limitations of the cloud and because OpenStack has reached its capacity. When OpenStack is not able to execute any workloads, we say that it fails. It can happen due to various reasons, for example as the result of the cumulative effect of ageing errors, that reduce OpenStack's capacity to 0, resource exhaustion such as physical space, virtual memory or network problems.

In addition to failure analysis, to identify trends in the ageing indicators we utilize the Mann-Kendall test at a 95\% confidence level, which is commonly used in software ageing studies~\cite{mann-kendall_2017, mann_kendall_confidence_95_2013}. To measure the strength of these trends, we employ Sen's slope method, which provides a reliable estimate of the rate of change in the ageing indicator. Through this analysis, we can quantify the manifestation of software ageing by evaluating the rate of change in the ageing indicators over time. Details for these methodologies are found in the Appendix~\ref{stats}.

Furthermore, we measure the total cumulative impact and ageing impact of software for each indicator. For an ageing indicator, for its starting value $v_0$ and its value before rejuvenation $v_b$ and its value after rejuvenation $v_r$, from the definition of ageing and rejuvenation, we define indicator's ageing $A$ and indicator's rejuvenation $R$ to be calculated as follows:
\[ A = v_b - v_0\] \[ R = v_b - v_r \]

By doing this we directly ascertain the total extent to which the cloud system's query execution impacts ageing indicators throughout a single scenario, and we obtain insights into the results of rejuvenation action.

\section{Evaluation}~\label{evaluation}
In this section, we undertake a comprehensive analysis of three scenarios, focusing initially on the influence of failures on OpenStack's ageing process and potential system failures. We particularly emphasize errors that occurred before the system reached a failed state, as post-failure errors did not significantly contribute to the ageing dynamics. Subsequently, we analyze the ageing patterns of workload duration and memory usage indicators across the scenarios. We compare these ageing patterns with the outcomes of the Mann-Kendall Test, Sen's slope, and our specific ageing and rejuvenation calculations. When constructing plots, a timeline that spans from 0 to 24 hours represents the ageing phase, the 25-hour mark corresponds to the rejuvenation phase, while the data at 26 hours signifies the post-rejuvenation phase. 

\subsection{Summarized Trend Analysis}
First, we present the outcomes of the trend evaluation using the Mann-Kendall test (MKT) and Sen's Slope for three key indicators: workload duration (WD), memory availability (MA), and swap usage. The application of MKT is feasible for a time series with a data volume equal to or exceeding 10, leading to the exclusion of the workload duration indicator for scenarios 5, 11, and 12 from this particular trend detection procedure~\cite[p.~211]{statistical_1987}. The detailed results of these calculations are provided in Table~\ref{table:trends}. Hereby, Sen's Slope is a measure that quantifies the rate of change of the target indicator over 1 hour. In the case of workload duration, it is expressed in seconds per hour, while for memory usage indicators, it is represented in gigabytes per hour. Cells in the table are colour-coded: green for indicating an upward trend, red for a downward trend, and yellow for the absence of any discernible trend.

\begin{table*}
\begin{center}
\begin{tabular}{ |>{\centering\arraybackslash}p{1.3cm}|>{\centering\arraybackslash}p{1.4cm}|>{\centering\arraybackslash}p{1.3cm}|>{\centering\arraybackslash}p{1.3cm}|>{\centering\arraybackslash}p{1.3cm}|>{\centering\arraybackslash}p{1.3cm}|>{\centering\arraybackslash}p{1.3cm}|>{\centering\arraybackslash}p{1.3cm}|  }
\hline
Scenario & Concur. & WD MKT & WD Slope & Swap MKT & Swap Slope & MA MKT & MA Slope  \\ 
\hline
\multicolumn{8}{|c|}{Multi-node} \\
\hline
  1 & 1 & \cellcolor{yellow!20}-1.81 & -0.03 & \cellcolor{green!20}6.45 & 0.005 & \cellcolor{red!20}-6.82 & -0.032 \\ 
  \hline
  2 & 2 & \cellcolor{green!20} 5.23 & 0.12 & \cellcolor{green!20}6.80 & 0.003 & \cellcolor{red!20}-4.59 & -0.033 \\ 
  \hline
  3 & 4 & \cellcolor{green!20} 4.99 & 0.23 & \cellcolor{green!20}6.82 & 0.011 & \cellcolor{red!20}-6.77 & -0.051\\ 
  \hline
  4 & 8 & \cellcolor{green!20} 5.18 & 0.56 & \cellcolor{green!20}5.95 & 0.036 & \cellcolor{red!20}-6.57 & -0.033\\ 
  \hline
  5 & 16 & - & - & \cellcolor{green!20}6.77 & 0.021 & \cellcolor{red!20}-4.44 & -0.014\\ 
  \hline
  6 & 64 & \cellcolor{red!20}-4.16 & -4.65 & \cellcolor{green!20}7.45 & 0.030 & \cellcolor{red!20}-4.45 & -0.011\\ 
  \hline
  
  \multicolumn{8}{|c|}{All-in-One} \\
\hline
  7 & 1 & \cellcolor{green!20}3.84 & 3.45 & \cellcolor{green!20}6.82 & 0.138 & \cellcolor{yellow!20}-0.97 & -0.004\\ 
  \hline
  8 & 2 & -0.32 & -0.46 & \cellcolor{green!20}6.77 & 0.143 & \cellcolor{yellow!20}1.51 & 0.008\\ 
  \hline
  9 & 4 & \cellcolor{green!20}6.33 & 14.61 & \cellcolor{green!20}6.82 & 0.148 & \cellcolor{green!20}2.95 & 0.016 \\ 
  \hline
  10 & 8 & \cellcolor{green!20}2.75 & 10.37 & \cellcolor{green!20}2.26 & 0.258 & \cellcolor{green!20}2.60 & 0.046\\ 
  \hline
  11 & 16 & - & - & \cellcolor{green!20}6.77 & 0.075 & \cellcolor{yellow!20}-0.22 & 0.002 \\ 
  \hline
  12 & 64 & - & - & \cellcolor{green!20}6.42 & 0.055 & \cellcolor{yellow!20}0.12 & 0.003\\ 
  \hline
\end{tabular}
\end{center}
\caption{Trend evaluation}
\label{table:trends}
\end{table*}

In Table~\ref{table:aging_summary}, we present the ageing (A) and rejuvenation (R) details for the identified ageing indicators. The indicators of ageing and rejuvenation represent measures that quantify the change of the target indicator over the first 24 hours and the 25-th hour, respectively. In the case of workload duration, it is expressed in seconds, while for memory usage indicators, it is represented in gigabytes. Notably, scenarios 4, 5, 11, and 12 are excluded from the analysis due to specific reasons. Scenario 4 lacked any rejuvenation process, while the other excluded scenarios did not generate more than 2 data points for the workload duration indicator. These specific scenarios will be examined in detail later in the analysis.

Apart from the trends we observed in the selected software ageing indicators, a notable trend emerged during our analysis of errors within the workload scenarios. Particularly in scenarios with concurrency 16 or 64, where a significant proportion of errors were associated with the exceeding of quotas for creating security groups. Notably, this specific error type accounted for over 96\% of the total errors in scenarios with a concurrency of 64. This frequent occurrence is attributed to the fact that the creation of security groups is the initial workload step limited by the capacity constraints of OpenStack. Consequently, this error type signifies an overloaded state of the OpenStack system rather than denoting critical errors.

Considering these observations, we decided to exclude the \texttt{Quota exceeded} errors for the \texttt{security group} OpenStack entity, but not for other ones, as they indicate OpenStack reduced capacity.

\subsection{Scenario Analysis}
\subsubsection{Multi-Node}
\begin{figure}[!t]
\centering
\includegraphics[width=0.5\textwidth]{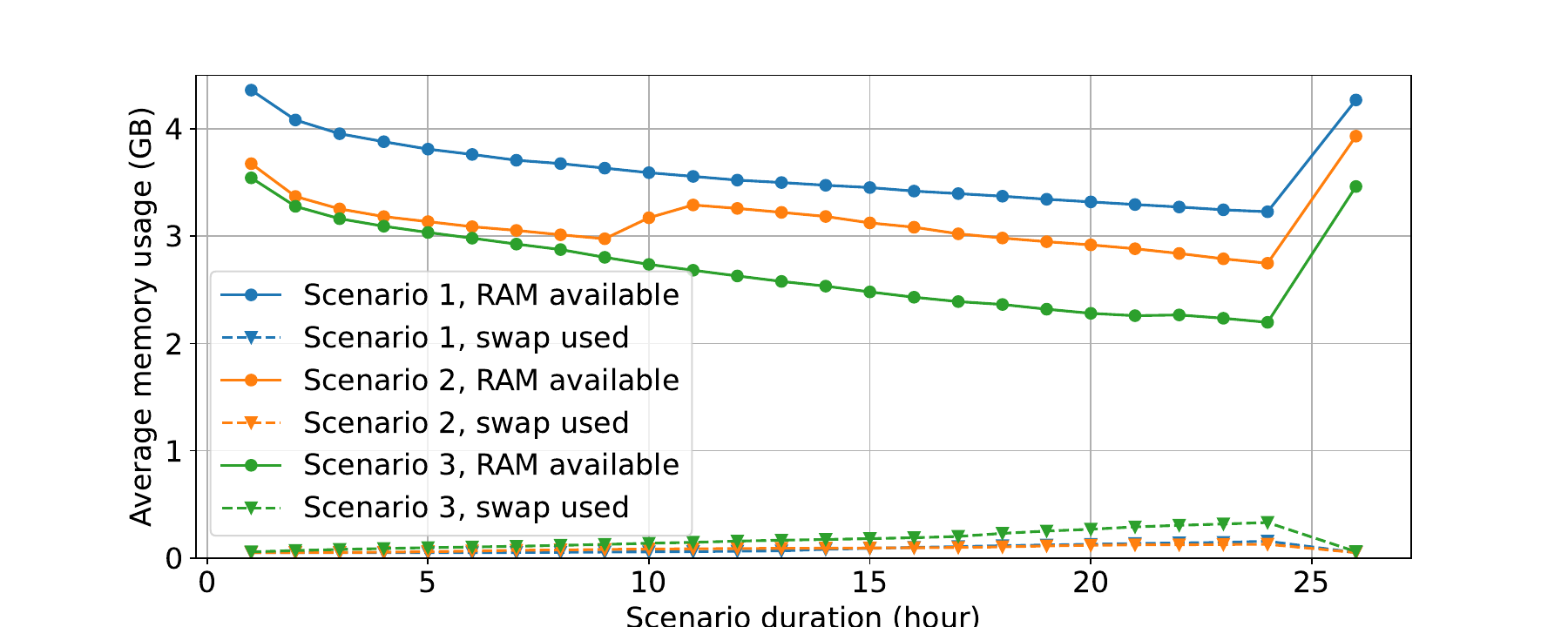}
\caption{Control node memory usage in scenarios 1, 2, 3.}
\label{figure/chapter_6/HA/HA_1_2_4_memory}
\end{figure}

Scenarios 1 and 3 did not report any errors, while scenario 2 encountered a single \texttt{Node is unreachable} error, which did not impact the OpenStack ageing process. Analysis of memory usage in Scenarios 1, 2, 3 (illustrated in Figure~\ref{figure/chapter_6/HA/HA_1_2_4_memory}) demonstrates a gradual decrease in available RAM during the ageing phase, with the amount of swap used gradually increasing. The MKT supports this observation, confirming that both ageing indicators displayed consistent ageing trends. 

In some cases, the ageing indicators rejuvenated more, than they have aged during the ageing phase. This can be explained by the fact, that we calculate average over 1 hour value. At the same time, the first hour of using OpenStack has large jumps and deviations in memory utilization. This is due to the so-called warm-up phase, during which lazy initiation of various objects within the system takes place. Because of this, two freshly installed OpenStacks can have a significant difference in memory usage during the first hour of use. But the main thing we see is that rejuvenation frees up all aged swap memory and a significant RAM portion.

In certain scenarios, our ageing indicators displayed more substantial rejuvenation than the ageing observed during the ageing phase. This difference can be attributed to the way we compute averages over 1-hour intervals. Additionally, during the initial hour of OpenStack usage, significant fluctuations and variations in memory utilization were apparent. These variations are commonly associated with the system's warm-up phase, during which the system's various objects are lazily initialized. Consequently, two newly installed OpenStack systems might exhibit notable differences in memory usage during the first hour of operation. Notably, rejuvenation released all the aged swap memory and a significant portion of RAM, as observed in our analysis.

\begin{figure}[!t]
\centering
\includegraphics[width=0.5\textwidth]{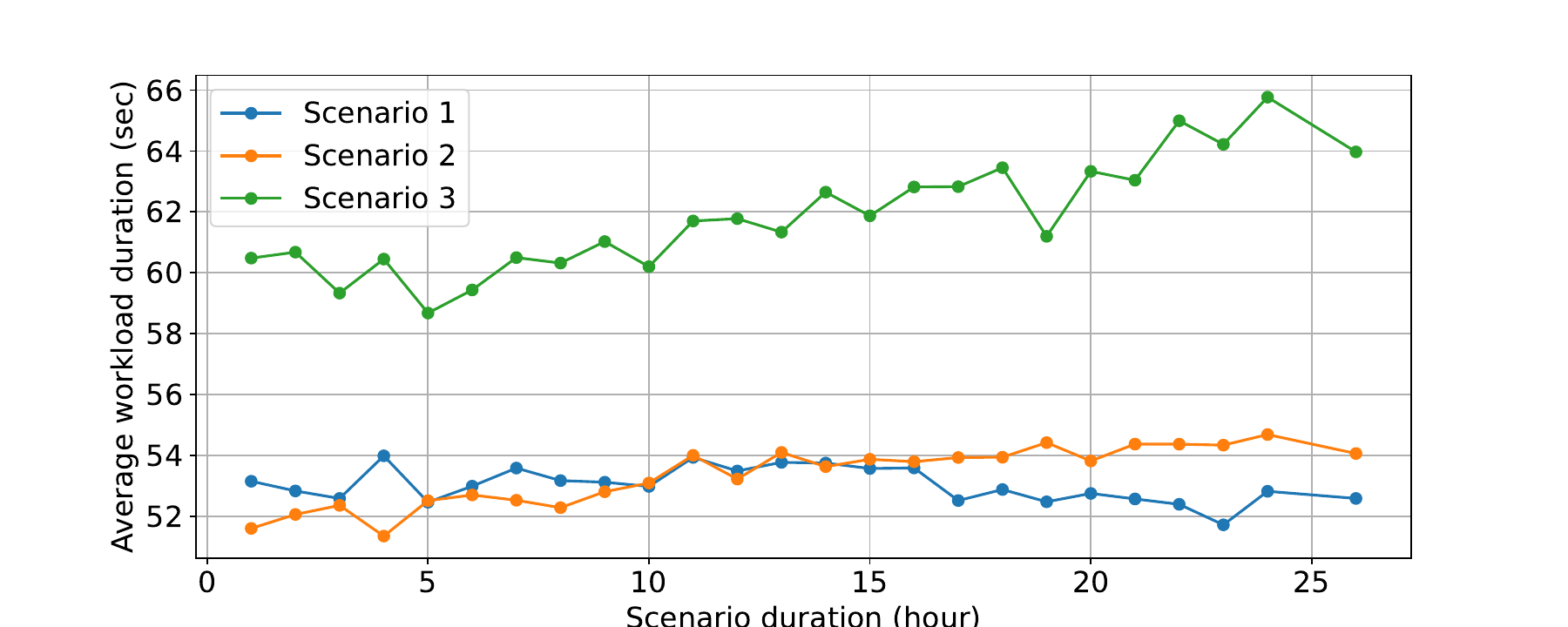}
\caption{Duration in scenarios 1, 2, 3.}
\label{figure/chapter_6/HA/HA_1_2_4_duration}
\end{figure}

Workload duration indicator changes during Scenarios 1, 2, and 3 are depicted in Figure~\ref{figure/chapter_6/HA/HA_1_2_4_duration}. MKT confirmed the presence of ageing trends only in scenarios 2 and 3.

\begin{figure}[!t]
\centering
\includegraphics[width=0.5\textwidth]{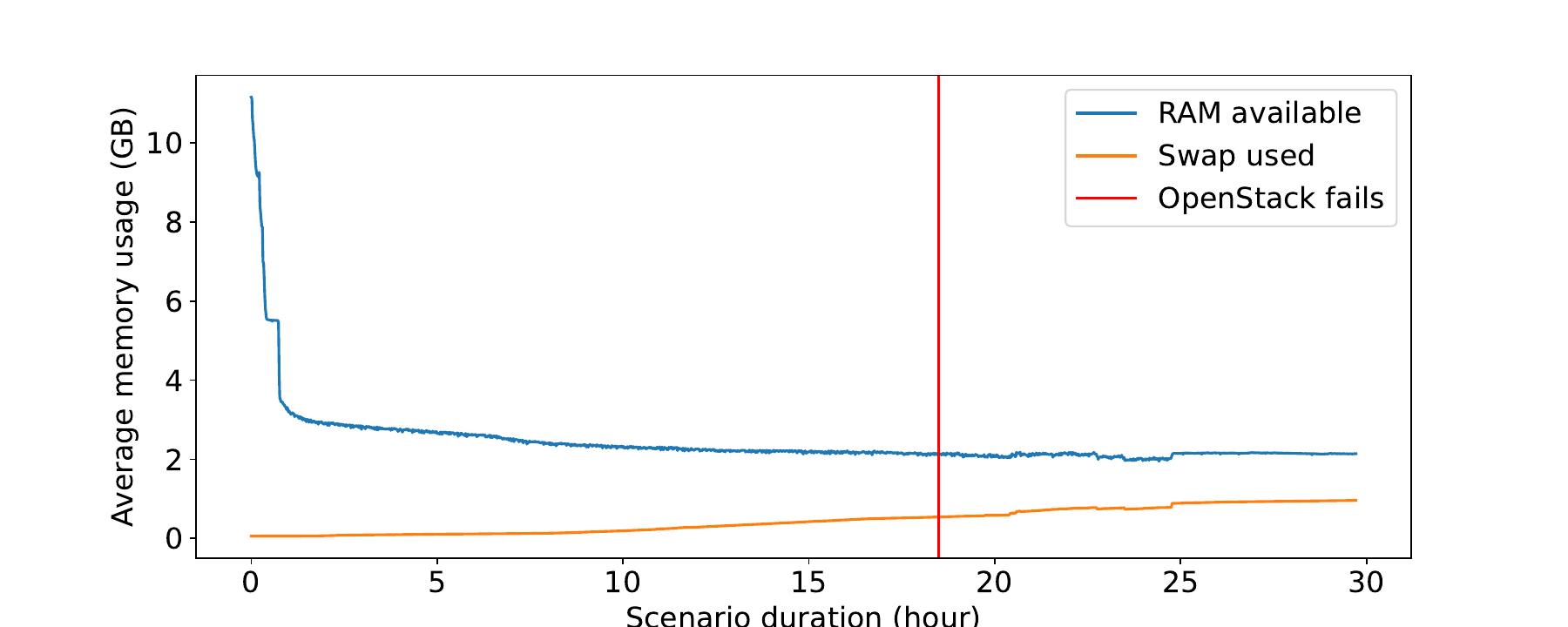}
\caption{Duration ageing in scenario 4.}
\label{figure/chapter_6/HA/HA_8_memory}
\end{figure}

We can see, that as expected, increased workload concurrency leads to faster software ageing. Notably, in scenario 3, which has the highest concurrency, all three indicators demonstrated the most prominent ageing effects, with the rejuvenation action restoring a substantial portion of the affected resources. Notably, only for workload duration, MKT could not identify the ageing trend in scenario 1, and in scenarios 2 and 3, WD rejuvenation was relatively small, reverting this indicator by less than 65\% of its aged value.

\begin{figure}[!t]
\centering
\includegraphics[width=0.5\textwidth]{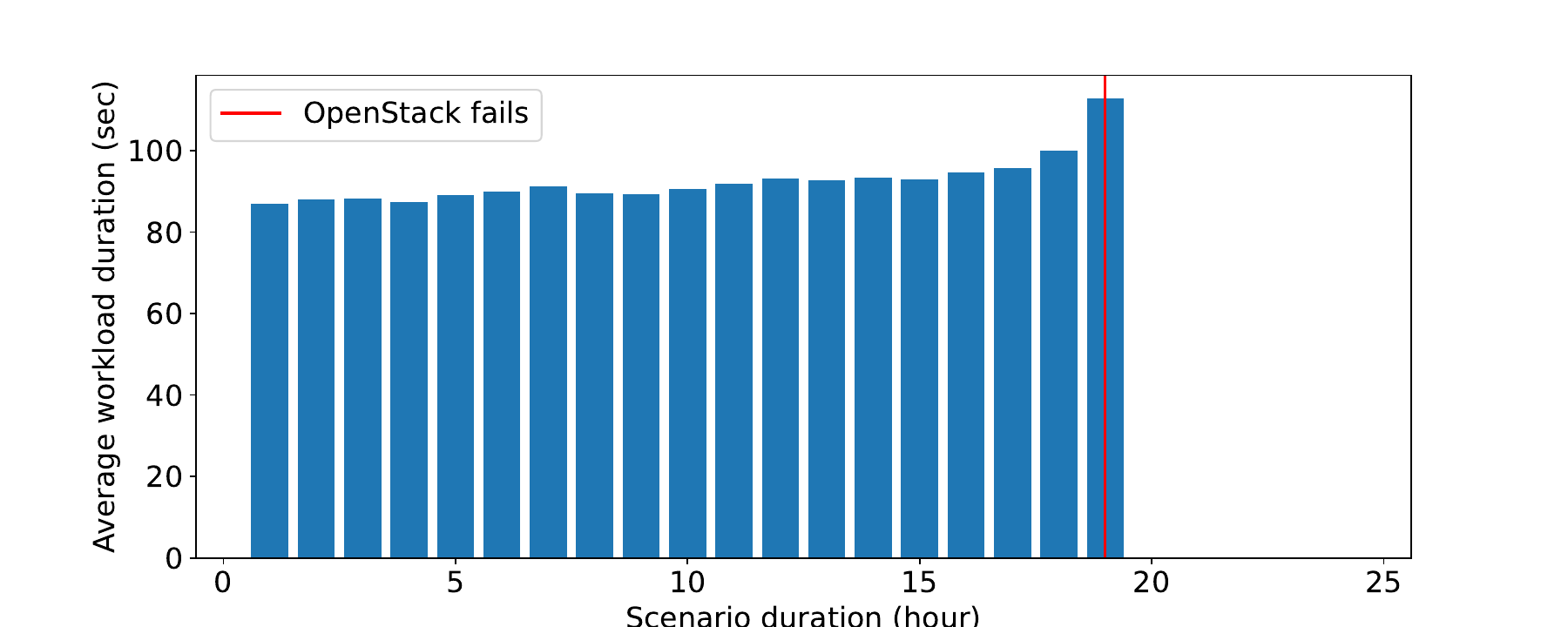}
\caption{Duration in scenario 4.}
\label{figure/chapter_6/HA/HA_8_duration}
\end{figure}

During Scenario 4, OpenStack encountered a critical failure, rendering it unable to execute further workloads. The failure occurred at approximately the 17.5-hour mark. At this critical point, the control node had 2.139 GB of available RAM and was utilizing 536 MB of swap memory. The average workload duration had increased by 26 seconds per hour. The ageing patterns of workload duration and memory consumption before the error state can be observed in Figure~\ref{figure/chapter_6/HA/HA_8_duration} and Figure~\ref{figure/chapter_6/HA/HA_8_memory}.

\begin{figure}[!t]
\centering
\includegraphics[width=0.5\textwidth]{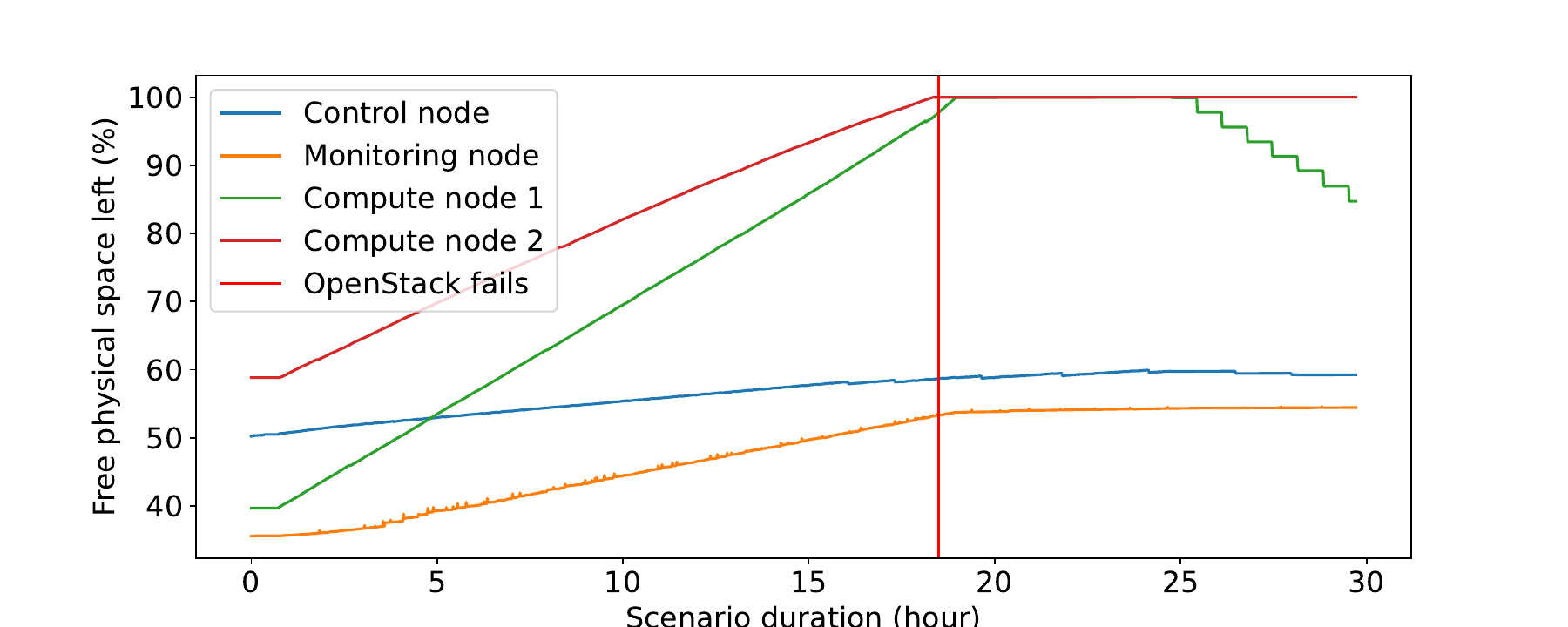}
\caption{Space usage on nodes in scenario 4.}
\label{figure/chapter_6/HA/HA_8_space}
\end{figure}

Following the exhaustion of physical space on both compute nodes, OpenStack entered a failed state, leading to the system failure without any preceding errors (Figure~\ref{figure/chapter_6/HA/HA_8_space}). This particular scenario represents the only instance where the depletion of OpenStack resources directly resulted in its failure. Subsequently, after the 24th hour of the scenario, Compute node number 1 initiated natural rejuvenation, gradually reclaiming physical space. In contrast, Compute node number 2 remained in an error state. Further investigation revealed unresponsiveness not only in the OpenStack environment but also in the Docker process on Compute node 2. Due to the inability of Kolla-ansible to operate on the affected machine, which hosted Compute node 2, we had to execute docker data pruning and restart the Docker process to restore the functionality of the affected system.

\begin{table*}
\begin{center}
\begin{tabular}
{|c|c|c|c|c|c|c|c|}
\hline
Scenario & Concur. & WD A & WD R & Swap A & Swap R & MA A & MA R  \\ 
\hline
\multicolumn{8}{|c|}{Multi-node} \\
\hline
  1 & 1 & -0.33& 0.24& 0.11& 0.11& -1.13& -1.04 \\ 
  \hline
  2 & 2 & 3.08& 0.62& 0.08& 0.08& -0.93& -1.18 \\ 
  \hline
  3 & 4 & 5.29& 1.80& 0.27& 0.27& -1.35& -1.27\\ 
  \hline
  6 & 64 & -44.26& -44.35& 0.43& 0.45& -0.77& -0.76\\ 
  \hline
  \multicolumn{8}{|c|}{All-in-One} \\
\hline
  7 & 1 & 75.24& 77.17& 3.24& 3.24& -1.12& -1.02\\ 
  \hline
  8 & 2 & 63.14& 57.00& 4.10& 4.00& -0.56& -0.58\\ 
  \hline
  9 & 4 & 420.54& 410.05& 3.67& 3.60& -1.31& -1.44\\ 
  \hline
  10 & 8 & 163.70& 194.04& 3.68& 3.69& 0.47& 0.64\\ 
  \hline
\end{tabular}
\end{center}
\caption{Ageing summary}
\label{table:aging_summary}
\end{table*}

Rally-OpenStack data analysis revealed that all workloads were successfully executed before the OpenStack failure, indicating that the corresponding entities created by these workloads were subsequently deleted. Notably, the system's depletion of physical space suggested that certain OpenStack atomic actions left residual files in the physical memory after the completion of the respective workloads. All workloads, that Rally was trying to execute past the failure point, were successfully executed until the Nova \texttt{Boot instance} action was reached. This implicated that atomic actions such as Cinders' \texttt{Create volume} and some of Nova's actions, including \texttt{Boot server}, \texttt{Attack volume}, and \texttt{Rebuild server}, potentially resulted in the persistence of leftover data. In pursuit of a resolution to this issue, an analysis of physical space usage by OpenStack containers in another All-in-One configuration after 20 hours of workload execution was done. Notably, we observed an unusual accumulation of files in the \texttt{nova\_compute/\_data/instances/\_base} folder within the Nova directory. Despite having only 2 failed instances and 2 running instances at the time of the investigation, the number of image instances in the specified folder totalled 1123. Each file associated with an instance was found to be 40MB in size, with total space usage of 44GB.

Subsequent investigation validated that the issue was indeed a result of a recognized software failure within the OpenStack system. According to the OpenStack documentation, under the default configuration, the Nova component uploads cache images on compute nodes but delays the cleaning process for unused images until after 24 hours, aligning precisely with the duration of our ageing phase. Throughout each scenario, OpenStack continued to generate these cache images without executing a comprehensive cleanup during the ageing phase. However, it was only in scenario 4 that OpenStack managed to execute a sufficient number of workloads to fill the physical disk space on the compute nodes.

\begin{figure}[!t]
\centering
\includegraphics[width=0.5\textwidth]{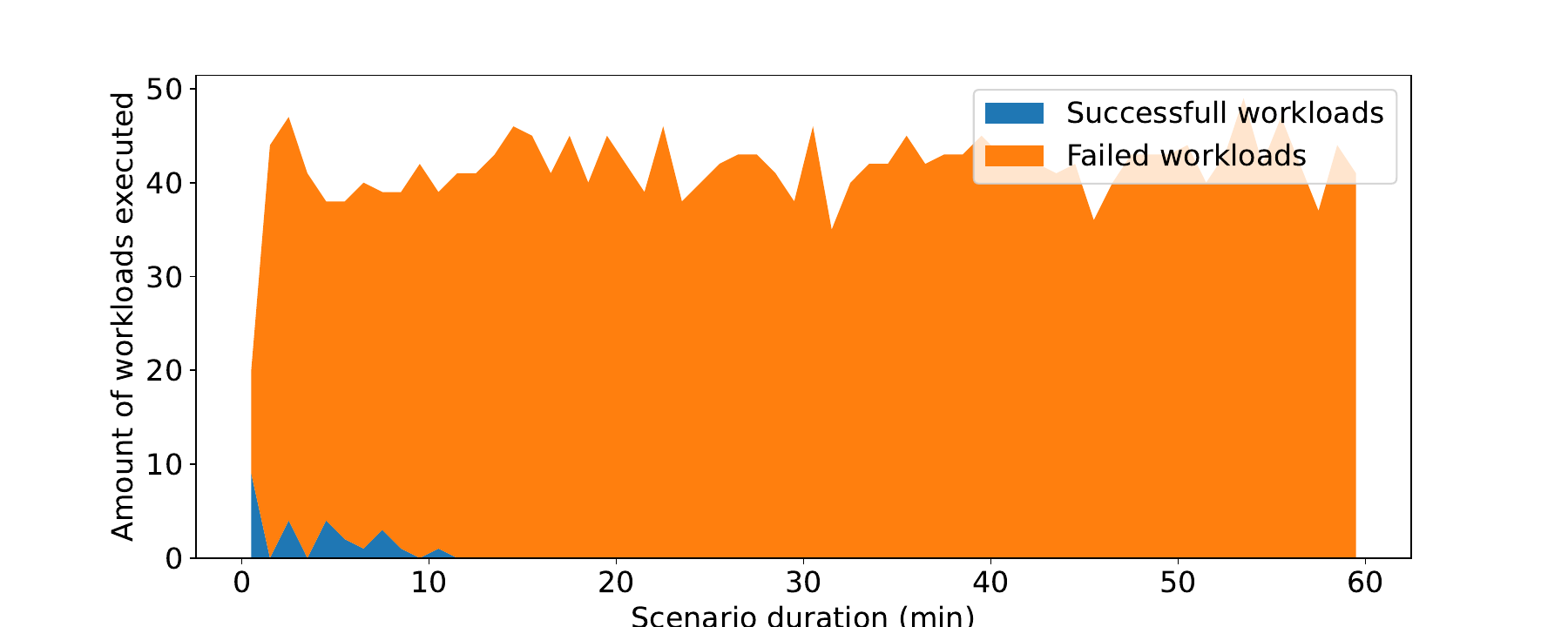}
\caption{Proportion of successful and failed workload executions during the first hour of scenario 5.}
\label{figure/chapter_6/HA/HA_16_first_hour}
\end{figure}

\begin{figure}[!t]
\centering
\includegraphics[width=0.5\textwidth]{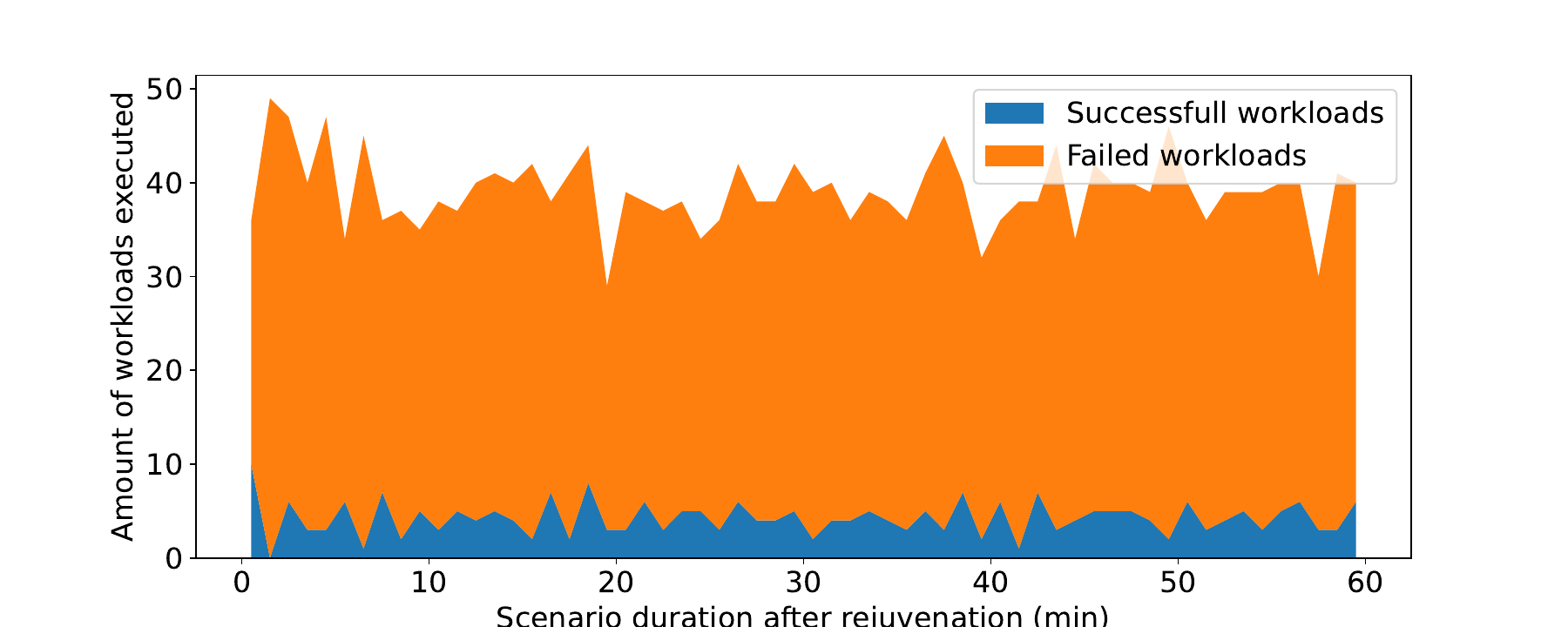}
\caption{Proportion of successful and failed workload executions after rejuvenation of scenario 5.}
\label{figure/chapter_6/HA/HA_16_after_rejuv}
\end{figure}

In Scenarios 5 and 6, OpenStack operated in an overloaded state from the outset, as the number of concurrent requests from Rally surpassed the established limits for various OpenStack entities. In scenario 5, OpenStack managed to execute only 25 workloads before experiencing a failure within the first 10 minutes due to exhaustion of its instance capacity, which generated \texttt{Create server} ageing errors (Figure~\ref{figure/chapter_6/HA/HA_16_first_hour}). During this period, certain workloads failed during the Nova \texttt{Boot instance} step, leaving residual instances in an \texttt{ERROR state}, consequently diminishing OpenStack's overall capacity with each additional virtual machine entity. However, after rejuvenation, OpenStack operated for an entire hour without entering the failed state (Figure~\ref{figure/chapter_6/HA/HA_16_after_rejuv}).

It is essential to acknowledge that in several scenarios, OpenStack became nonfunctional following the initial deployment or rejuvenation, necessitating a secondary redeployment action for the cloud, either before the ageing phase or as part of the rejuvenation process. This highlights the manifestation of the system in different states following consecutive deployments. In scenario 5 the system showed improved performance after the rejuvenation, then at the beginning of the ageing phase. This phenomenon can be described as computing inconsistency. It emphasizes the significance of conducting multiple repetitions of experiments, particularly when engaging in complex activities such as deploying a cloud solution. It underscores the importance of accounting for variability in system behaviour, even when executing identical computing operations, such as the installation of a fresh OpenStack system.

\begin{figure}[!t]
\centering
\includegraphics[width=0.5\textwidth]{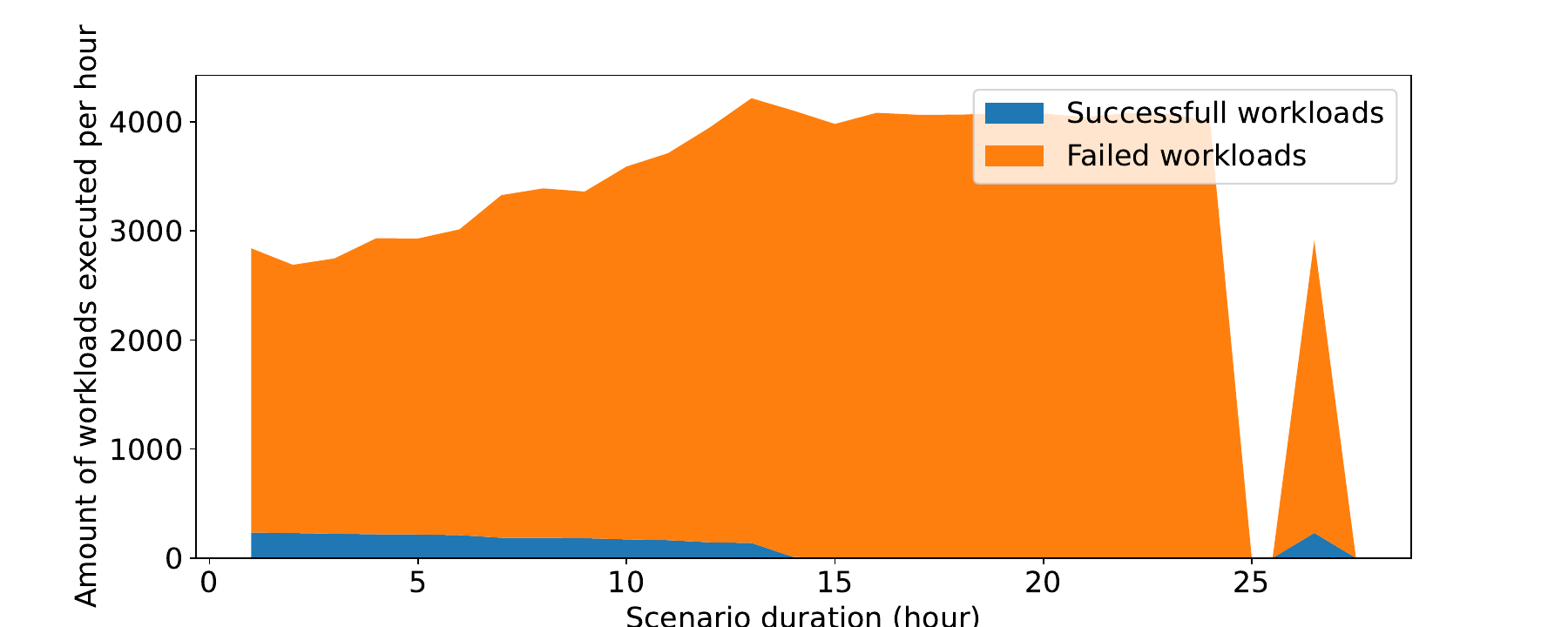}
\caption{Proportion of successful and failed workload executions in scenario 6.}
\label{figure/chapter_6/HA/HA_64_succ_fail}
\end{figure}

During scenario 6, the system successfully executed workloads for the initial 14 hours, continuously operating in an overloaded state, as depicted in Figure~\ref{figure/chapter_6/HA/HA_64_succ_fail}. Throughout this period, the quantity of successfully executed workloads per hour consistently diminished, while the average workload duration remained relatively stable for the first 6 hours. Subsequently, the average workload duration began to decrease over the remaining 7 hours until OpenStack entered an error state.

\begin{figure}[!t]
\centering
\includegraphics[width=0.5\textwidth]{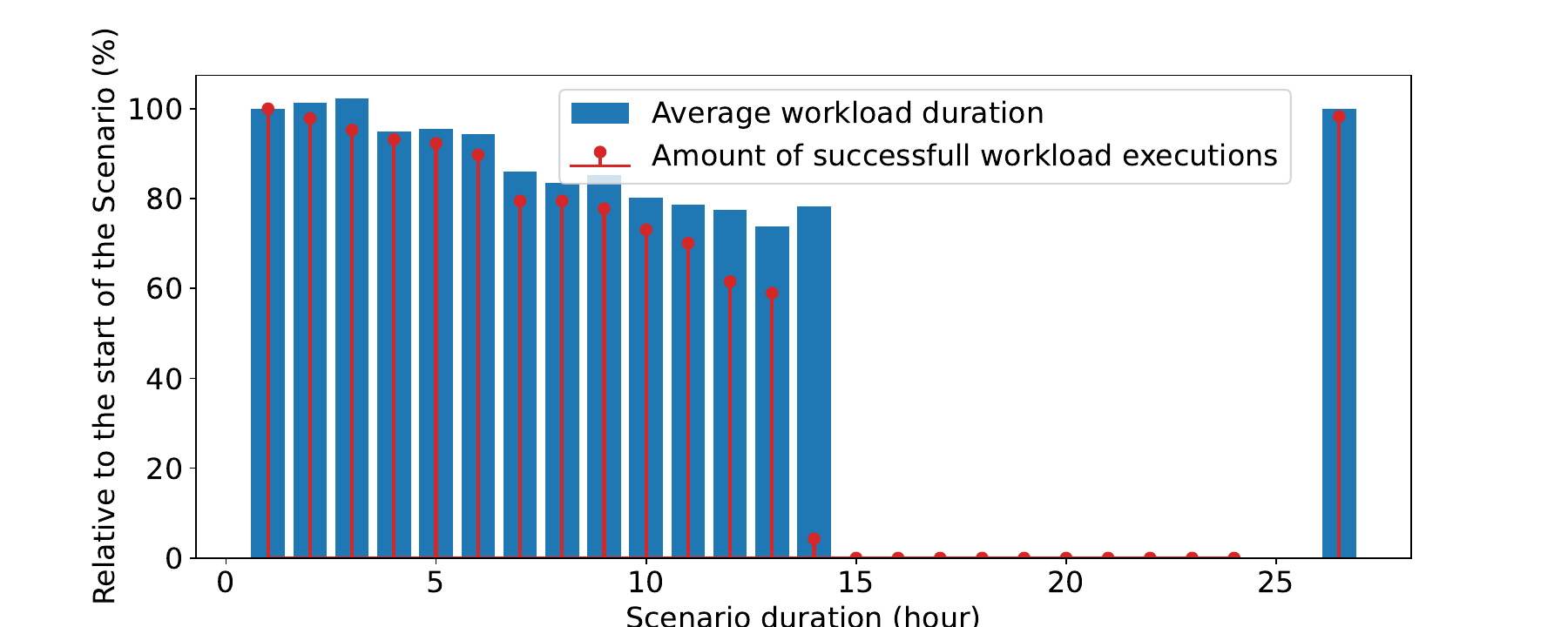}
\caption{Correlation between average workload duration and amount of successful workload executions in scenario 6.}
\label{figure/chapter_6/HA/HA_64_dur_succ}
\end{figure}

In this scenario, the duration of the workload as an ageing indicator contradicted the expected ageing pattern. As it can be seen in Figure~\ref{figure/chapter_6/HA/HA_64_dur_succ}, instead of increasing, workload duration had a downward trend, which is confirmed by MKT. At the same time, the rate of successful workload executions was decreasing as well. While conclusive evidence is unavailable, it is plausible that very high initial values of workload duration were caused by the fact, that OverStack resources were being concurrently used by all ten concurrently running workloads. However, as OpenStack's capacity gradually diminished due to errors, the number of concurrently fully executable workloads decreased, resulting in fewer workloads sharing the same resources. It is important to note that such a downward trend in the workload duration indicator was not observed in other scenarios, as errors leading to OpenStack failure typically transpired within a single hour rather than over 13 hours. 

\begin{figure}[!t]
\centering
\includegraphics[width=0.48\textwidth]{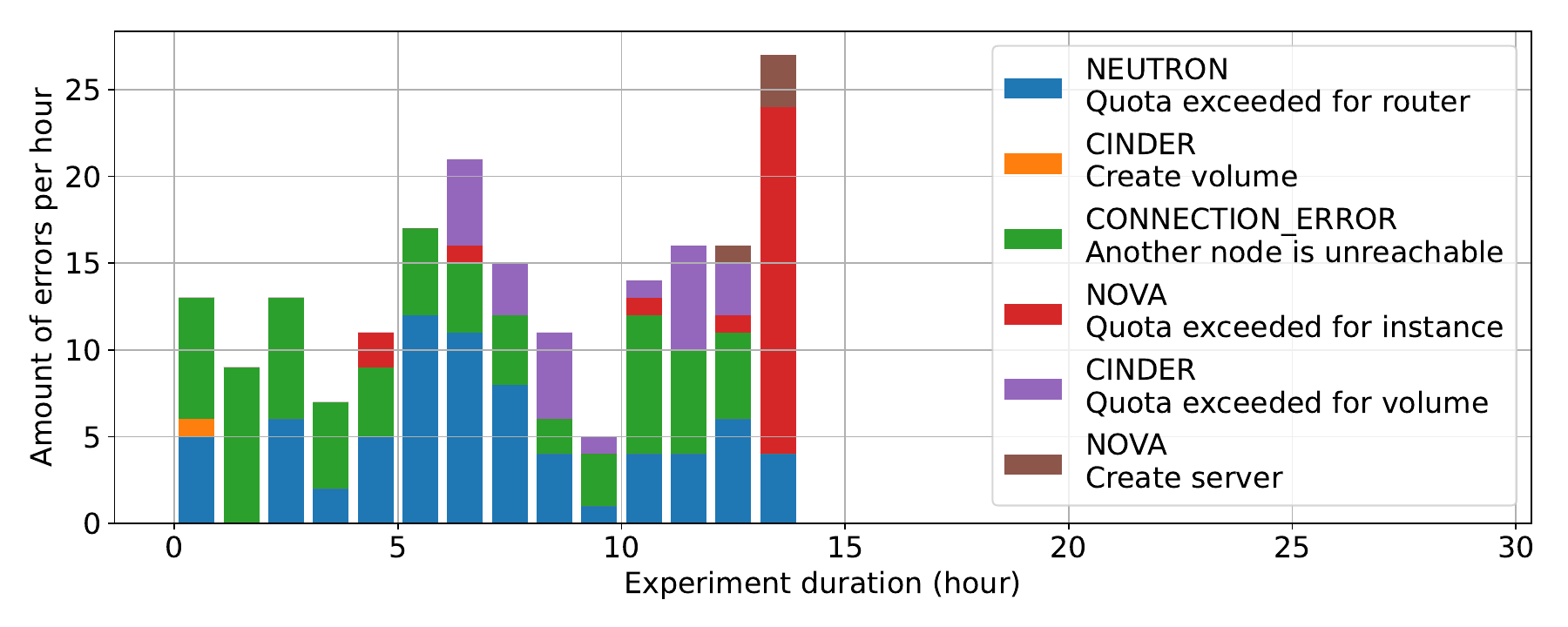}
\caption{Errors in scenario 6.}
\label{figure/chapter_6/failures/HA_64_errors}
\end{figure}

Error distribution in scenario 6 can be observed in Figure~\ref{figure/chapter_6/failures/HA_64_errors}. 
As can be seen, multiple errors occur before the OpenStack failure point. From this list \texttt{Quota exceeded for security group} is excluded, as it was already mentioned in this chapter. Thus, we expected the first \texttt{Quota exceeded} error to occur for the entity, for which the very first ageing error occurs.

This was not the case, because error \texttt{Another node is unreachable} occurred persistently from the start of the experiment until the point of OpenStack failure. Since this error is associated with networking issues, its ageing nature depends on the phase of the workload during which it occurs. Notably, out of the 69 occurrences of the error "Another node is unreachable," 44 incidents were noted during the "Create user" workload step. Given that this step is the initial phase of the workload, it is probable that it did not leave any discernible OpenStack entities. However, complete assurance regarding this matter cannot be provided, as it is plausible that the OpenStack user was indeed created, yet no explicit evidence of its creation was found.
Furthermore, 20 additional errors were distributed among the actions of "Create role," "Delete role," "Add role," "Revoke role," and "Delete user." While these errors did not directly influence OpenStack's capacity, they might have left certain entities that could potentially contribute to the ageing effect on OpenStack's resource usage and workload duration.
Three instances of the error "Another node is unreachable" were observed during the "Attach volume" step, while the final two occurrences were noted during the "Detach volume" workload steps. These errors can be classified as ageing errors since they transpired after the creation of OpenStack entities, manifesting as early as the first and third hours of the experiment, consequently potentially reducing OpenStack's operational capacity.
Notably, a single instance of the "Volume has error status" ageing error was recorded during the first hour of the scenario. However, this was followed by multiple occurrences of the "Quota exceeded for volume" error, which emerged after at least five hours had elapsed. Similarly, the "Quota exceeded for instance" error manifested as early as the fourth hour, while the "Instance has ERROR status" error only appeared at the 14th hour of the experiment. These findings support the idea that the "Another node is unreachable" error significantly contributed to OpenStack ageing during the initial five-hour period.

The accumulation of residual entities within OpenStack, particularly those that diminish its capacity, contributed to a gradual decline in its overall performance. This manifested as a reduced capability to effectively execute workloads, ultimately leading to its failure.

Scenarios 5 and 6 exhibited distinct failure points and varied error distributions, which resulted in different impacts on the system's ageing. However, comparing the workload duration indicator between these two scenarios is challenging due to the limited number of workloads executed in scenario 5. Yet, even the incomparability in terms of workload duration data between scenarios 5 and 6 signifies that ageing manifested differently in these two.

\begin{figure}[!t]
\centering
\includegraphics[width=0.48\textwidth]{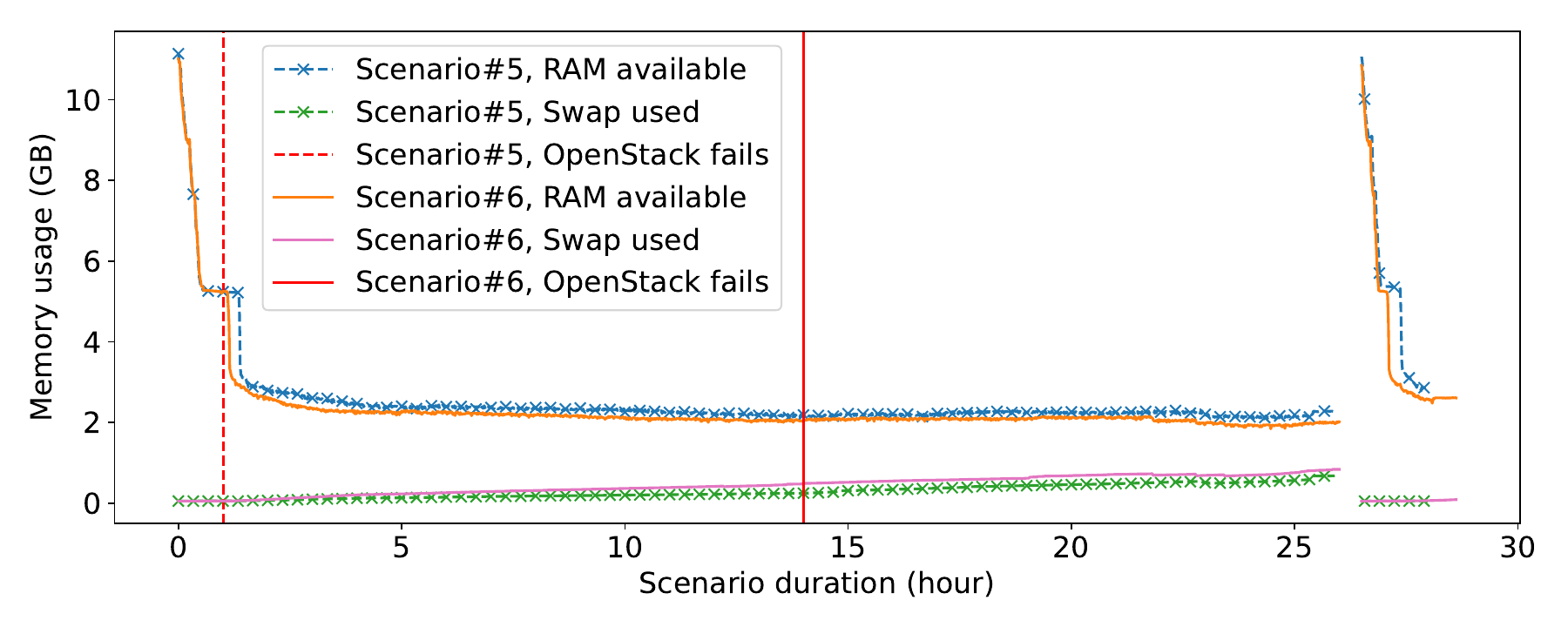}
\caption{Errors in scenario 6.}
\label{figure/chapter_6/HA/HA_16_64_memory}
\end{figure}

Simultaneously, a comparison of memory usage ageing indicators between Scenario 5 and 6 reveals strikingly similar patterns (Figure~\ref{figure/chapter_6/HA/HA_16_64_memory}), which do not markedly differ from the memory usage pattern observed in other multi-node configuration scenarios.

MKT indicates the same trends in memory usage indicators and only Sen's slope shows an anomaly, which demonstrates lower values in scenarios 5 and 6 in contrast to other multi-node scenarios, despite the increased concurrency of workload executions. Consequently, these findings suggest that workload duration, particularly when combined with the number of successful workloads per hour, offers more crucial insights into the effects of software ageing than memory usage ageing indicators in these particular scenarios.

\subsubsection{All-in-One}
With the All-in-One configuration, OpenStack performed workloads without entering a failed state in scenarios 7, 8, and 9. In scenario 7, a single \texttt{Rebuild server error} occurred, yet it did not seem to impact OpenStack. Similarly, in scenario 8, there were two \texttt{Create server} errors in the fourth hour of the experiment and seven \texttt{External network unreachable} errors, all of which did not appear to affect OpenStack ageing. On the other hand, scenario 9 did not encounter any errors. Regarding workload duration, the MKT identified the presence of an ageing trend in scenarios 7 and 9, which was supported by Sen's Slope and our calculations of the workload duration ageing (same as in Table \ref{table:aging_summary}).

\begin{figure}[!t]
\centering
\includegraphics[width=0.48\textwidth]{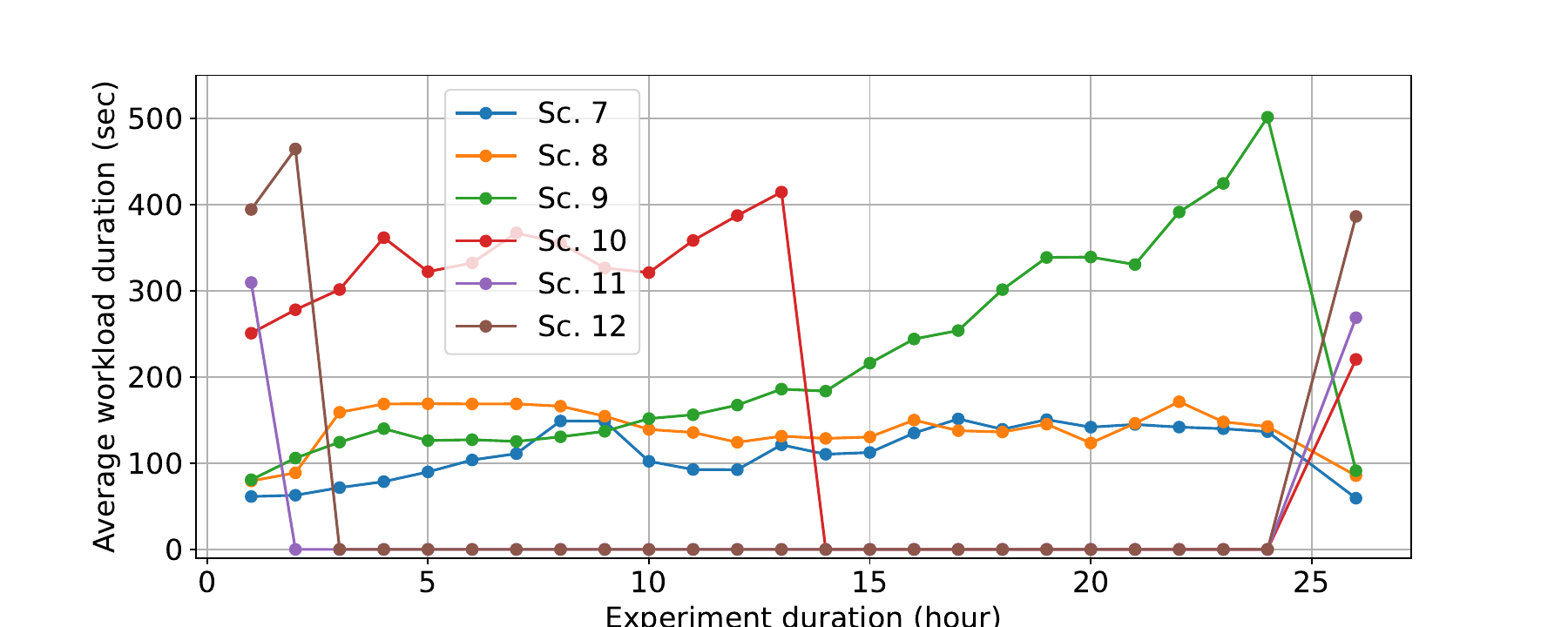}
\caption{Scenarios 7-12, average successful workload duration.}
\label{figure/chapter_6/AiO/AiO_duration}
\end{figure}

MKT indicated the absence of ageing in scenario 8. However, this assertion is challenging to make due to the doubling of workload duration during the first two hours, followed by stabilization at a certain level with slight fluctuations, as illustrated in Figure~\ref{figure/chapter_6/AiO/AiO_duration}. Workload duration experienced an increase of 63 seconds per average workload execution time during the ageing phase, followed by a rejuvenation of 57 seconds per average workload execution time in the rejuvenation phase. Moreover, the potential intersection of ageing and overload on this ageing indicator further complicates the definitive assessment of software ageing in this scenario. This particular situation emphasizes the need to reevaluate our methodology, as a step for further work, particularly the necessity of implementing the wait phase from the SWARE methodology. This additional step would allow OpenStack to internally rejuvenate itself in the absence of a load, potentially altering the ageing pattern in this scenario.

Simultaneously, we observe an exceptionally strong trend in scenario 9. This is demonstrated by both Sen's slope and our calculations of ageing and rejuvenation for this ageing indicator, as well as in Figure~\ref{figure/chapter_6/AiO/AiO_duration}, where the indicator displays almost monotonic growth. This observation is consistent with our understanding of the influence of configuration and concurrency on software ageing. 

\begin{figure}[!t]
\centering
\includegraphics[width=0.48\textwidth]{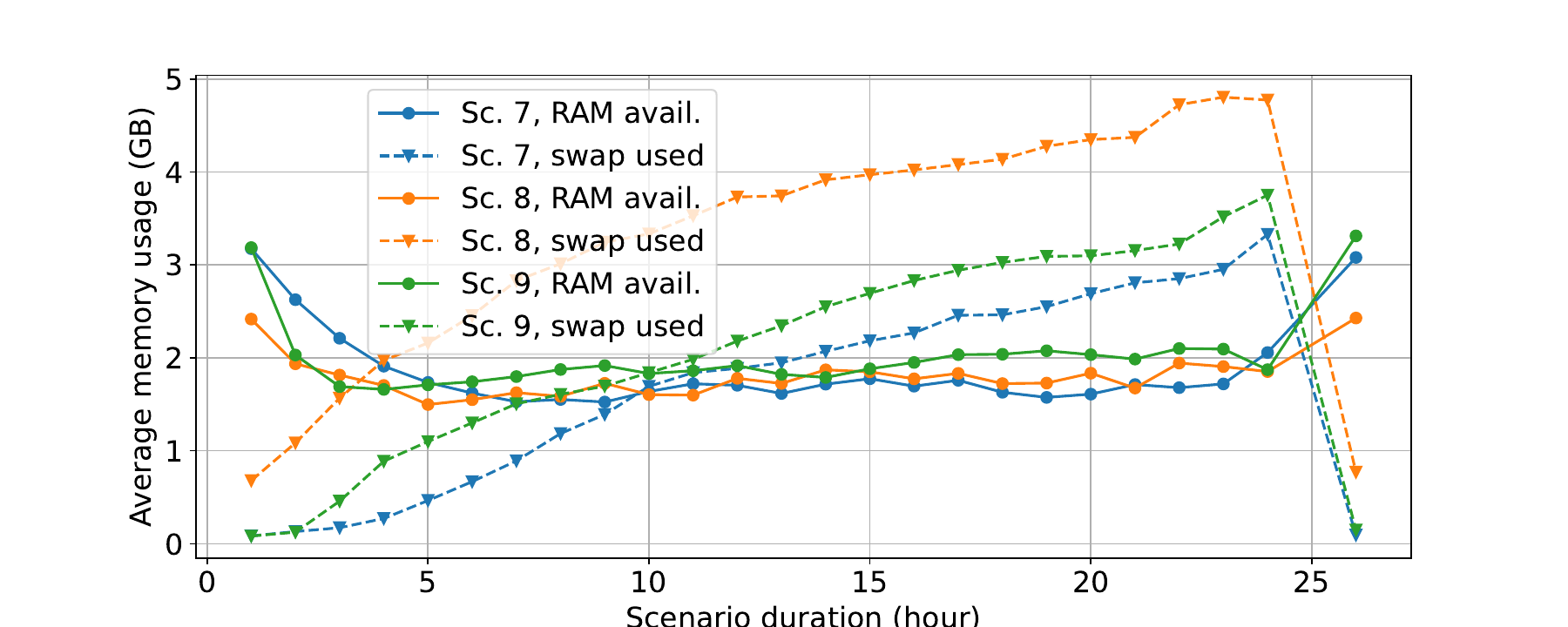}
\caption{Memory usage in scenarios 7, 8, 9.}
\label{figure/chapter_6/AiO/AiO_1_2_4_memory}
\end{figure}

Memory consumption indicators for scenarios 7, 8, and 9 exhibit a pattern distinct from their counterparts in the multi-node configuration on Figure~\ref{figure/chapter_6/AiO/AiO_1_2_4_memory}. MKT does not detect an ageing trend in scenarios 7 and 8 for the Memory Available indicator. In scenario 9, according to our calculations, this indicator has aged by -1.31 GB and was rejuvenated by -1.44 GB. However, MKT indicates an opposite trend, suggesting an increase in Memory Available during scenario 9. Upon closer examination of Figure~\ref{figure/chapter_6/AiO/AiO_1_2_4_memory}, it becomes evident that the Memory Available indicator experienced a significant initial surge, followed by stabilization at a certain value. This may explain why MKT detected an upward trend, conflicting with our analytical analysis and understanding of ageing as a cumulative effect. At the same time, Swap ageing exhibits a much stronger trend, as indicated by both trend evaluation and our calculations of ageing/rejuvenation for the indicators. These data can be attributed to the fact that OpenStack does not have much memory available following its deployment, resulting in small fluctuations throughout a single scenario. Meanwhile, Swap used remains non-zero from the first hour, experiencing a notably faster growth rate compared to the multi-node scenario, only to be rejuvenated to its initial values.

\begin{figure}[!t]
\centering
\includegraphics[width=0.48\textwidth]{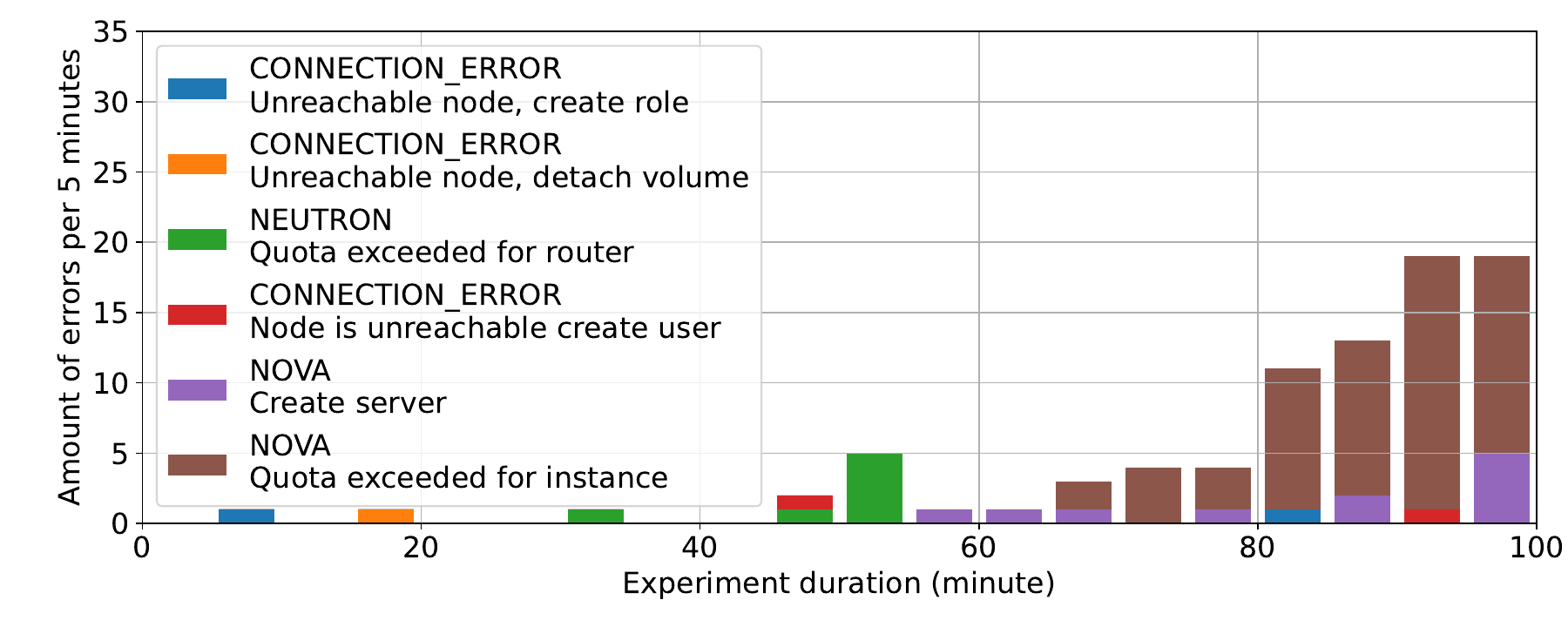}
\caption{Errors in Scenario 12.}
\label{figure/chapter_6/failures/AiO_64_errors}
\end{figure}

\begin{figure}[!t]
\centering
\includegraphics[width=0.48\textwidth]{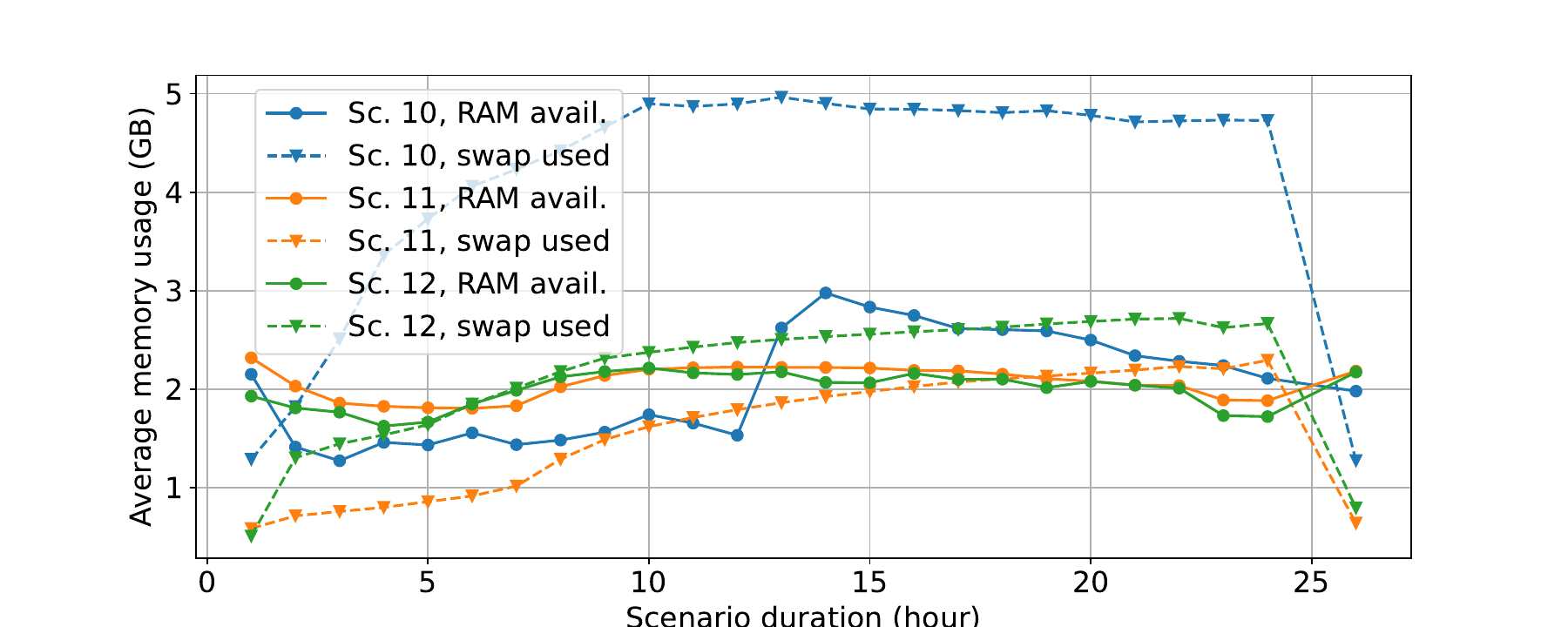}
\caption{Memory usage.}
\label{figure/chapter_6/AiO/AiO_8_16_64_memory}
\end{figure}

Such a difference in the behaviour of ageing indicators across different configurations underscores the importance of considering not only the presence or strength of a trend but also the initial value of the indicators, their limitations, and their connection with other ageing indicators. In scenario 10, a single \texttt{Rebuild server} error occurred in the fourth hour of the scenario, followed by a sequence of 9 \texttt{Create server} errors, ultimately failing OpenStack. In scenario 11, a single \texttt{Create volume} error occurred at the beginning of the scenario, and subsequently, 10 non-sequential \texttt{Create server} errors led the cloud to a failed state. In scenario 12, during the first hour, the following failures occurred: 1 \texttt{Create server} error, a single \texttt{Another node is unreachable} error for each create role, user, and detach volume step, and multiple \texttt{Quota exceeded for router} errors (Figure~\ref{figure/chapter_6/failures/AiO_64_errors}). OpenStack was still able to execute some workloads until it failed with a sequence of 9 \texttt{Create server} errors during the second hour of workload, ultimately reducing its capacity to 0. We cannot analyze the workload duration indicator for scenarios 11 and 12, but scenario 10, provided the expected results: both MKT and our calculations of indicator ageing and rejuvenation indicated the presence of an ageing trend. In scenario 10, the trend intensity is lower than in scenario 9, despite scenario 10 having higher concurrency. However, this can be explained by the fact that OpenStack failed in scenario 10, which altered the expected outcome of the workload duration ageing indicator.
On Figure~\ref{figure/chapter_6/AiO/AiO_8_16_64_memory} we can observe changes in memory usage.

On Figure~\ref{figure/chapter_6/AiO/AiO_8_16_64_memory}, we can observe the changes in memory usage ageing indicators. For the Memory Available indicator, MKT indicated the presence of a trend only for scenario 10, and once again a contradictory one – the amount of Memory Available improved throughout the experiment. However, Figure~\ref{figure/chapter_6/AiO/AiO_8_16_64_memory} shows that when OpenStack entered the failed state, a large portion of memory was released, which likely indicates the result of internal rejuvenation. For scenarios 11 and 12, no trend was detected, which can be explained by the initially low value of memory available. The Swap Used indicator shows the existence of an ageing trend in all three scenarios, yet the ageing in scenarios where OpenStack entered the failed state early is noticeably slowed down.

\subsection{Findings}

In this study, errors occurred in all scenarios, except for 1, 3, and 9. Despite this, OpenStack reached a failed state only in Scenarios 4, 5, 6, 10, 11, and 12, which corresponds to scenarios with a concurrency of 8 and higher. Moreover, the entry into the failed state occurred through different error patterns, demonstrating the significance of conducting ageing experiments with various concurrency parameters.

One of the primary objectives of the cloud is to provide fault tolerance, enabling cloud users to continue using virtualized solutions even when errors arise within the cloud. Continuing the experiment during errors until the cloud loses the ability to perform workloads provides valuable insights into how errors impact the cloud's fault tolerance. Incorporating minimal native rejuvenation into the workload allows for an even better understanding of this process.

It is evident that, in general, a higher concurrency value accelerates ageing trends in workload duration and swap usage indicators. Additionally, we conclude that a higher concurrency value speeds up the trends in memory available only if there is RAM that the host operating system is ready to allocate to the cloud. If there is no such reserve, then the ageing trends in swap are accelerated.

Furthermore, our findings have provided clear evidence that an all-in-one configuration exhibits lower initial performance, and more intense ageing, and only scenario 6 for the multi-node configuration demonstrated a consistently improving ageing indicator during the ageing process. These observations underscore the importance of incorporating different configurations in cloud ageing studies.

Considering that scenarios 4, 5, 6, 10, 11, and 12 turned out to be non-standard, a closer examination of the table reveals that memory usage does not provide sufficient information about the ageing of the system. Furthermore, we observe that commonly used memory usage indicators in ageing studies have little correlation with the workload duration indicator, and in some scenarios, they even develop in opposite directions. This suggests that their connection with the ageing process varies; they react differently to it, indicating that they identify the ageing process from different perspectives.

In some scenarios, the time to execute one workload increased by hundreds of per cent during the ageing process, indicating a significant deterioration in cloud elasticity as well as user experience. Information about OpenStack response time and workload execution frequency can be utilized to directly calculate how much time is wasted due to ageing effects, while information about memory usage can be used for the same purpose only indirectly - by taking into account the knowledge that swap is slower than RAM and has a direct impact on request response time. Thus, response time proves to be more useful for calculating the cost efficiency of rejuvenation, if there is no task of preventing memory resource deficiency.

Thus, we see that request response time is a reliable support as an OpenStack ageing indicator, especially when used in combination with workload execution frequency and error analysis.

\subsection{Effectiveness Evaluation}
In the process of trend analysis, it became apparent that our scenario definition should be extended by two additional phases: an additional wait phase and an additional ageing phase. The inclusion of the wait phase, which was initially introduced by De Melo et al., is not directly feasible due to the inability to track the request response time on an idle system \cite{sware_2017_approach_rejuvenation}. However, a reasonable compromise could involve introducing a wait phase, during which indicators are not monitored, followed by another ageing phase which will provide data, necessary for the analysis of ageing. This approach would offer insight into the extent to which the system rejuvenates itself during the wait phase, leading to a better understanding of the balance between an overloaded and ageing system.

Throughout many scenarios, OpenStack entered a failed state because we have implemented only a single-point fault tolerance at the workload definition level, which does not sufficiently meet the requirements of modern cloud systems. Modern cloud solutions are very robust and have a very large resource reserve. For better fault tolerance implementation, an additional procedure for cleaning up leftover OpenStack entities should be added. This can be done as the extending of our workload clean-up mechanism, which can attempt to delete error entities that were created as the result of error steps themselves. Or it can be implemented at the scenario level, which attempt to delete residual error entities after a failed workload.

The period following OpenStack's failure up to the point of rejuvenation did not yield meaningful data. This highlights that continuing to execute workloads after OpenStack's failure, resulting from reaching capacity 0, does not contribute significant research insights. Therefore, triggering rejuvenation autonomously at an earlier stage when the system identifies OpenStack's failure would be a more efficient approach compared to the current methodology that waits for a specific temporal point.

The detection of trends in the indicators using the Mann-Kendall test concerning the confidence threshold of 95\% coincided with our assessment and understanding of trends in 25 out of 27 cases. Only in workload duration in scenario 8 and Memory Available in scenario 9, the MKT result was in doubt. Sen's Slope also provides a good estimation of ageing trends. Its values closely match our ageing indicator calculations in Table \ref{table:aging_summary}, but only for the swap indicator. For example, in scenario 3, Sen's Slope gives us an ageing rate of 0.011 GB per hour, which amounts to 0.264 over 24 hours. Our ageing indicator calculation yielded 0.27. In scenario 9, Sen's Slope indicates an ageing rate of 0.148 GB per hour, which amounts to 3.55 over 24 hours, whereas our indicator gave 3.67 GB. However, for other indicators, Sen's Slope readings may differ from our calculations. The usage of both these two calculation methods and trend visual analysis has proven useful because it has provided us with a more comprehensive picture for trend intensity analysis, where one of the aforementioned indicators would show a one-sided perspective.

The result of the Mann-Kendall test, along with the encountered errors, showed that the selected ageing indicators complement each other well in the analysis of software ageing. Out of 12 scenarios, only two scenarios, specifically scenarios 2 and 3, showed complete correlation between the results of the Mann-Kendall test on different indicators. In other scenarios, the selected ageing indicators either proved to be not sufficiently expressive when used in isolation or even diverged from each other in different directions, which confirms our initial hyptohesis about the importance of diversifying ageing indicators and proves the combination of chosen age indicators and ageing evaluation methodology to be effective.

\section{Related Work}
The foundation for developing an effective methodology and designing efficient experiments lies in an extensive body of literature that addresses the study of software ageing. Key works in this domain provide insights into the tasks involved in studying software ageing, explain the methodologies employed, and outline the evolving trends within the field. Notably, literature focusing on the role of rejuvenation and the design of rejuvenation strategies proves crucial for shaping our approach. Additionally, works directly addressing software ageing in the context of cloud computing are of paramount importance. These works delve into the specifics of studying ageing phenomena within the specifics of cloud environments, offering valuable insights into industry-wide trends, methodologies, and preferred data types in this domain. Furthermore, exploring works specifically dedicated to software ageing and performance in the OpenStack system is essential. This exploration allows us to comprehend the behaviour of OpenStack indicators. It lets us learn methodologies for analyzing ageing within OpenStack, compare them with broader cloud computing practices, and evaluate the scope of tasks they address concerning the overarching goals of studying software ageing.

\subsection{Software Ageing Studies}
Cotroneo et al. analyzed 71 papers from important conferences and journals in the field of software ageing and rejuvenation of long-running software projects \cite{aging_rejuv_where_we_are_2011}. They found that the majority of studies in this field focus on developing predictive models for determining optimal rejuvenation times, often relying on memory and performance reduction indicators. Their findings suggested a need for further research to understand more complex signs of software ageing and address potential errors in calculation methods. They revealed a trend of lack of experimentation on real systems, partly because 41\% of the papers that have been analyzed were model-based studies that validate their approach by numerical examples. In our study, we address these concerns by proposing a methodology for conducting an experimental study of software ageing inside a cloud and by investigating the usefulness of alternative ageing indicators.

Alonso et al. conducted a comparative experimental study on six software rejuvenation techniques aimed at mitigating software ageing effects and evaluating their performance overhead \cite{rejuvcomparative}. Their study emphasizes the critical impact of rejuvenation granularity on its effectiveness, giving us information to determine the more appropriate level of rejuvenation for our study. The results also indicate that the level of granularity in rejuvenation techniques impacts performance overhead, with virtualization introducing additional memory fragmentation challenges, leading to comprehensive guidelines for designing rejuvenation scheduling algorithms and selecting appropriate mechanisms. Their work helps us better understand the impact of rejuvenation and its granularity on our experiment data.

\subsection{Software Ageing in Cloud Computing}
Pietrantuono and Russo did a study about software ageing and rejuvenation in cloud systems by examining a set of 105 papers from various digital libraries \cite{aging_rejuv_survey_2020}. They showed, that the cloud ageing phenomena have been analyzed almost the same number of times by model-based as by measurement-based techniques, with a non-negligible number of studies addressing the analysis by hybrid approaches. They highlight the increasing interest in this domain, with a focus on rejuvenation solutions for both measurement-based and model-based ageing analysis methods.
They showed that for measurement-based ageing assessment, time series analysis is the most used technique to analyze the data and some of the frequently used methods are the Mann-Kendall test and Sen's Slope, which we adopt as our main age evaluation technique. They also demonstrate the dominance of memory indicators as an ageing metric. According to their study, request response time as an ageing indicator is not a commonly used metric. Their study confirms the importance of our research in addressing and exploring the utilization of additional metrics alongside memory utilization, highlighting the need for a more comprehensive understanding of software ageing in cloud environments.

Oliveira et al. presented an experimental evaluation of software ageing effects in dockerd daemon \cite{docker_aging_memory_2020}. Dockerd is the process that is responsible for the management of Docker containers and consequently is supposed to support long-running systems, which makes it a potential target of software ageing. They performed a simple workload of creation and deletion of containers, which was executed on a single machine. During the design of the methodology and data analysis, they followed the aforementioned SWARE approach. As a result of their study, they provided evidence of software ageing happening inside dockerd process up to a catastrophic failure, when dockerd became unresponsive, and some brute-force actions were needed to recover the system.

Torquato et al. evaluated software ageing effects on the Docker platform in a cloud computing environment \cite{docker_aging_2019}. As a testbed, a single machine with Docker service was used. To speed up the ageing process two scenarios with different workloads were conducted. The first one in the form of a cycle of creating, restarting, and deleting containers to simulate a cloud computing environment similar to that used in cloud infrastructure provisioning. Another scenario was defined as a cycle of repeated interaction with 5 existing Docker containers to observe the ageing effect inside a container on a long-running setup. CPU, disk usage, and memory usage were chosen as metrics of interest. They observed in the first scenario memory leaks in the NetworkManager process so that after some time the free memory supply became insufficient for Docker to instantiate new containers, affecting system reliability. Their second scenario showed that with long-running containers, Docker process containers can suffer from increased memory usage. Both studies of software ageing in docker demonstrate its vulnerability to ageing effects in situations of both short- and long-living containers. 

\subsection{OpenStack Performance and Software Ageing}
Melo et al. measured the utilization of hardware and software resources after applying a stressful workload on an OpenStack all-in-one environment \cite{openstack_aging_mem_usage_component_melo_2017}. Their workload consists of repeated instantiations and terminations of virtual machines. For environment monitoring, various Linux utilities were used that collected information about the general consumption of the hardware resources as well as about resources' consumption of the processes related to the OpenStack and the service provision. During the study, they used a concept of trend analysis to predict resource utilization through time with the help of time series. Their experiment ended with a system crush in the form of a workload failure after 3 days of executing workloads. Authors do not specify whether they have tried to continue workload executions after the first failure, and the type of failure that has occurred. We want to extend their methodology of software ageing evaluation by including some degree of fault tolerance by continuing workload execution even with failures present. 

Pflanzner et a. analyzed the performance of OpenStack by conducting a series of experiments on it~\cite{openstack_performance_rally_2016}. They have defined a set of 9 scenarios with various workloads, targeting different cloud services, that were executed both in concurrent and non-concurrent manner. Workloads consist of different interactions with OpenStacks modules, such as Compute, Network, Image, and Storage. To execute workloads and collect experiment results, a module named Rally-OpenStack was used. The authors of this study planned to use custom Rally scripts to introduce custom workloads but encountered technical difficulties, which led them to use Python OpenStack API. Our workload includes a majority of actions, included in scenarios of this study, and are also executed by Rally-OpenStack, but are defined in a form of a single Rally script. Authors of this study pointed out, that concurrent execution of workloads generally slows down request execution time, which we expect to see in the results of our study as well. In some scenarios, they also experienced failures of an irregular nature both on the level of an OpenStack module and at the hardware level, which also happens in our study due to the obscure nature of the ageing process.

Cotroneo et al. did an empirical study of software failures in OpenStack~\cite{openstack_failures_2019}. They have analyzed the impact of module failures in terms of fail-stop behaviour, failure detection through logging, and failure propagation across components. The analysis points out that most of the failures are not detected. Moreover, many of these failures can silently propagate over time and through components of the cloud management system, which calls for more thorough run-time checks and fault containment. In their study, they used fault injection to simulate a situation, when a software failure happens in OpenStack. They demonstrated that failures are distributed across different OpenStack modules and also across different user requests to those modules, which means that a wide range of modules and the actions that the modules perform have to be a part of the investigation for a complete assessment of software ageing in OpenStack. Propagation of failures between different OpenStack modules and the long-lasting effect of failures, even when the initial failure causes have been addressed means, that when one service ages, making it more prone to software failures, it also increases the risk of other services experiencing failures. It leads to a potential increase in the rate of software ageing in those services as well. Accordingly, this confirms the importance of studying the ageing effect not only at the level of OpenStack subsystems but also of the OpenStack system as a whole. Cotroneo et al. also confirmed, that some failures that were studied, were not logged or notified to the user. It confirms that the information about the most severe and direct impact of software ageing, failures, potentially disregards a certain amount of data, which increases the importance of other ageing indicators that can be measured without loss of information, such as response time or memory usage.

Upon thorough analysis of the literature, it becomes evident that despite the significance of request response time as a key characteristic of OpenStack performance, it remains underutilized as an ageing indicator, which underscores the relevance of this research. Through a review of experimental study methodologies and the examination of gathered data, we have encountered instances illustrating the design of experiments aimed at imposing stress on OpenStack. These works have also contributed valuable insights into how software ageing evaluation is typically conducted, which we use as a foundational reference for shaping our own evaluation methodology.

Additionally, our examination has brought to light a notable gap in information related to the analysis of software failures in the context of OpenStack. Recognizing this gap, we are poised to address and investigate software failures comprehensively as part of our evaluation process, contributing to a more comprehensive understanding of software ageing effects in the OpenStack environment.

\section{Conclusion}~\label{conclusion}
This paper aimed to investigate software ageing and rejuvenation in the context of cloud systems, with a particular focus on OpenStack. The comprehensive study encompassed an in-depth analysis of software ageing patterns, failures, and chosen rejuvenation strategy, utilizing a series of purposefully designed experiments. Throughout the research, various methodologies were employed to evaluate the effects of software ageing, with specific emphasis on request response time and memory usage.

Through an in-depth evaluation of the collected data, we made several insights into the different software ageing indicators and their change due to failures. Our analysis highlighted various factors contributing to the divergences between the capability of memory utilisation and response time to detect the software ageing effect and showed the crucial ageing trends associated with distinct configurations and levels of concurrency. We observe that commonly used memory usage indicators have little correlation with the workload duration indicator, and in some scenarios, they even develop in opposite directions. This suggests that their connection with the ageing process varies; indicating that they identify the ageing process from different perspectives. Memory usage does not always provide sufficient information about the ageing of the system, prompting complementary analysis of multiple ageing factors. In some scenarios, the time to execute one workload increased by more than 100\% during the ageing process, indicating a significant deterioration in cloud elasticity. For example, information about OpenStack response time and workload execution frequency can be utilized to directly calculate how much time is wasted due to ageing effects, while information about memory usage can be used for the same purpose only indirectly. Thus, response time proves to be more useful for calculating the cost efficiency of rejuvenation. Thus, we see that request response time is a reliable support as an OpenStack ageing indicator, especially when used in combination with workload execution frequency and error analysis, extending the observations from prior works.

The study underscored the significance of employing an effective ageing evaluation methodology, as demonstrated by the depth of the comparative analysis of different evaluation metrics. Overall, this paper sheds light on the intricacies of software ageing in distributed systems, particularly within the OpenStack framework. It serves as a foundation for future research endeavours, offering valuable insights into failure analysis, the role of ageing indicators and ageing evaluation methodologies. The findings of this study provide a basis for further investigations into optimizing the performance and longevity of distributed systems. We released the web application, which was used to conclude scenario execution, as a free open-source software \footnote{https://github.com/Ydjeen/openstack\_testbed} that can be used by anyone interested in it.

\bibliographystyle{IEEEtran}
\bibliography{IEEEabrv,referencesValid.bib}

\begin{thebibliography}{10}
\providecommand{\url}[1]{#1}
\csname url@samestyle\endcsname
\providecommand{\newblock}{\relax}
\providecommand{\bibinfo}[2]{#2}
\providecommand{\BIBentrySTDinterwordspacing}{\spaceskip=0pt\relax}
\providecommand{\BIBentryALTinterwordstretchfactor}{4}
\providecommand{\BIBentryALTinterwordspacing}{\spaceskip=\fontdimen2\font plus
\BIBentryALTinterwordstretchfactor\fontdimen3\font minus
  \fontdimen4\font\relax}
\providecommand{\BIBforeignlanguage}[2]{{%
\expandafter\ifx\csname l@#1\endcsname\relax
\typeout{** WARNING: IEEEtran.bst: No hyphenation pattern has been}%
\typeout{** loaded for the language `#1'. Using the pattern for}%
\typeout{** the default language instead.}%
\else
\language=\csname l@#1\endcsname
\fi
#2}}
\providecommand{\BIBdecl}{\relax}
\BIBdecl

\bibitem{cloudcomputing_2020}
A.~Sunyaev, ``Cloud computing,'' \emph{Internet Computing: Principles of
  Distributed Systems and Emerging Internet-Based Technologies}, pp. 195--236,
  2020.

\bibitem{cloud_risks_2009}
W.~Kim, ``Cloud computing: Today and tomorrow.'' \emph{J. Object Technol.},
  vol.~8, no.~1, pp. 65--72, 2009.

\bibitem{cloud_risks_2014}
R.~Latif, H.~Abbas, S.~Assar, and Q.~Ali, ``Cloud computing risk assessment: a
  systematic literature review,'' \emph{Future Information Technology:
  FutureTech 2013}, pp. 285--295, 2014.

\bibitem{dependabilitydefinition2004}
A.~Avi\v{z}ienis, J.-C. Laprie, B.~Randell, and C.~Landwehr, ``Basic concepts
  and taxonomy of dependable and secure computing,'' \emph{IEEE Transactions on
  Dependable and Secure Computing}, vol.~1, no.~1, pp. 11--33, 2004.

\bibitem{Notoro2021}
P.~Notaro, J.~Cardoso, and M.~Gerndt, ``A survey of aiops methods for failure
  management,'' \emph{ACM Trans. Intell. Syst. Technol.}, vol.~12, 2021.

\bibitem{soft_aging_fundamentals_2008}
M.~Grottke, R.~Matias, and K.~S. Trivedi, ``The fundamentals of software
  aging,'' in \emph{2008 IEEE International conference on software reliability
  engineering workshops (ISSRE Wksp)}.\hskip 1em plus 0.5em minus 0.4em\relax
  Ieee, 2008, pp. 1--6.

\bibitem{software_rejuvenation_1995}
Y.~Huang, C.~Kintala, N.~Kolettis, and N.~D. Fulton, ``Software rejuvenation:
  Analysis, module and applications,'' in \emph{Twenty-fifth international
  symposium on fault-tolerant computing. Digest of papers}.\hskip 1em plus
  0.5em minus 0.4em\relax IEEE, 1995, pp. 381--390.

\bibitem{aging_rejuv_survey_2020}
R.~Pietrantuono and S.~Russo, ``A survey on software aging and rejuvenation in
  the cloud,'' \emph{Software Quality Journal}, vol.~28, no.~1, pp. 7--38,
  2020.

\bibitem{aging_rejuv_where_we_are_2011}
D.~Cotroneo, R.~Natella, R.~Pietrantuono, and S.~Russo, ``Software aging and
  rejuvenation: Where we are and where we are going,'' in \emph{2011 IEEE Third
  International Workshop on Software Aging and Rejuvenation}.\hskip 1em plus
  0.5em minus 0.4em\relax IEEE, 2011, pp. 1--6.

\bibitem{aging_eucalyptus_2011_memory}
J.~Araujo, R.~Matos, P.~Maciel, R.~Matias, and I.~Beicker, ``Experimental
  evaluation of software aging effects on the eucalyptus cloud computing
  infrastructure,'' in \emph{Proceedings of the middleware 2011 industry track
  workshop}, 2011, pp. 1--7.

\bibitem{aging_response_hosted_app_2021}
E.~Andrade, R.~Pietrantuono, F.~Machida, and D.~Cotroneo, ``A comparative
  analysis of software aging in image classifiers on cloud and edge,''
  \emph{IEEE Transactions on Dependable and Secure Computing}, 2021.

\bibitem{response_time_app_aws_2010}
M.~Alhamad, T.~Dillon, C.~Wu, and E.~Chang, ``Response time for cloud computing
  providers,'' in \emph{Proceedings of the 12th International Conference on
  Information Integration and Web-based Applications \& Services}, 2010, pp.
  603--606.

\bibitem{openstack_performance_rally_2016}
T.~Pflanzner, R.~Tornyai, B.~Gibizer, A.~Schmidt, and A.~Kertesz, ``Performance
  analysis of an openstack private cloud,'' 2016.

\bibitem{openstack_nasa_popular_aws_2013}
S.~Yadav, ``Comparative study on open source software for cloud computing
  platform: Eucalyptus, openstack and opennebula,'' \emph{International Journal
  Of Engineering And Science}, vol.~3, no.~10, pp. 51--54, 2013.

\bibitem{openstack_architecture_2014}
T.~Rosado and J.~Bernardino, ``An overview of openstack architecture,'' in
  \emph{Proceedings of the 18th International Database Engineering \&
  Applications Symposium}, 2014, pp. 366--367.

\bibitem{kolla-ansible_testbed_2021}
X.~Liu, T.~Embuldeniya, Z.~Liu, Z.~Li, M.~A. Garcia, C.~McAllister, and
  D.~Schwieger, ``An initial exploration of multi-node open cloud
  infrastructure testbed,'' in \emph{Proceedings of the Conference on
  Information Systems Applied Research ISSN}, vol. 2167, 2021, p. 1508.

\bibitem{kolla-ansible_testbed_2020}
C.~Vitucci, T.~Cucinotta, R.~Mancini, L.~Abeni \emph{et~al.}, ``Implementation
  and deployment of a server at the edge using openstack components,'' in
  \emph{Proceedings of the 19th International Conference on Networks (ICN
  2020)}.\hskip 1em plus 0.5em minus 0.4em\relax IARIA, 2020.

\bibitem{kolla_rally_2017}
R.-A. Cherrueau, D.~Pertin, A.~Simonet, A.~Lebre, and M.~Simonin, ``Toward a
  holistic framework for conducting scientific evaluations of openstack,'' in
  \emph{2017 17th IEEE/ACM International Symposium on Cluster, Cloud and Grid
  Computing (CCGRID)}.\hskip 1em plus 0.5em minus 0.4em\relax IEEE, 2017, pp.
  544--548.

\bibitem{software_aging_1994}
D.~L. Parnas, ``Software aging,'' in \emph{Proceedings of 16th International
  Conference on Software Engineering}.\hskip 1em plus 0.5em minus 0.4em\relax
  IEEE, 1994, pp. 279--287.

\bibitem{software_rejuv_hybrid_2013}
J.~Zhao, Y.~Wang, G.~Ning, K.~S. Trivedi, R.~Matias~Jr, and K.-Y. Cai, ``A
  comprehensive approach to optimal software rejuvenation,'' \emph{Performance
  Evaluation}, vol.~70, no.~11, pp. 917--933, 2013.

\bibitem{docker_aging_2019}
M.~Torquato and M.~Vieira, ``An experimental study of software aging and
  rejuvenation in dockerd,'' in \emph{2019 15th European Dependable Computing
  Conference (EDCC)}.\hskip 1em plus 0.5em minus 0.4em\relax IEEE, 2019, pp.
  1--6.

\bibitem{docker_aging_2021}
F.~Oliveira, J.~Araujo, R.~Matos, and P.~Maciel, ``Software aging in
  container-based virtualization: an experimental analysis on docker
  platform,'' in \emph{2021 16th Iberian Conference on Information Systems and
  Technologies (CISTI)}.\hskip 1em plus 0.5em minus 0.4em\relax IEEE, 2021, pp.
  1--7.

\bibitem{docker_aging_memory_2020}
F.~Oliveira, J.~Araujo, R.~Matos, L.~Lins, A.~Rodrigues, and P.~Maciel,
  ``Experimental evaluation of software aging effects in a container-based
  virtualization platform,'' in \emph{2020 IEEE International Conference on
  Systems, Man, and Cybernetics (SMC)}.\hskip 1em plus 0.5em minus 0.4em\relax
  IEEE, 2020, pp. 414--419.

\bibitem{software_aging_app_in_cloud_2021}
R.~Matos, J.~Araujo, V.~Alves, and P.~Maciel, ``Characterization of software
  aging effects in elastic storage mechanisms for private clouds,'' in
  \emph{2012 IEEE 23rd International Symposium on Software Reliability
  Engineering Workshops}.\hskip 1em plus 0.5em minus 0.4em\relax IEEE, 2012,
  pp. 293--298.

\bibitem{rejuv_redundancy_2012}
D.~S. Kim, S.~M. Lee, J.-H. Jung, T.~H. Kim, S.~Lee, and J.~S. Park,
  ``Reliability and availability analysis for an on board computer in a
  satellite system using standby redundancy and rejuvenation,'' \emph{Journal
  of mechanical science and technology}, vol.~26, pp. 2059--2063, 2012.

\bibitem{sware_2017_approach_rejuvenation}
M.~D.~T. de~Melo, J.~Araujo, I.~Umesh, and P.~R.~M. Maciel, ``Sware: an
  approach to support software aging and rejuvenation experiments,''
  \emph{Journal on Advances in Theoretical and Applied Informatics}, vol.~3,
  no.~1, pp. 31--38, 2017.

\bibitem{guedes2019availability}
E.~A.~C. GUEDES, ``Availability and capacity modeling for virtual network
  functions based on redundancy and rejuvenation supported through live
  migration,'' 2019.

\bibitem{openstack_aging_10_hours}
H.~Meng, Y.~Shi, Y.~Qu, J.~Li, and J.~Liu, ``Arima-based aging prediction
  method for cloud server system,'' in \emph{IOP Conference Series: Materials
  Science and Engineering}, vol. 1043, no.~2.\hskip 1em plus 0.5em minus
  0.4em\relax IOP Publishing, 2021, p. 022021.

\bibitem{araujo2011software}
J.~Araujo, R.~Matos, P.~Maciel, and R.~Matias, ``Software aging issues on the
  eucalyptus cloud computing infrastructure,'' in \emph{2011 IEEE international
  conference on systems, man, and cybernetics}.\hskip 1em plus 0.5em minus
  0.4em\relax IEEE, 2011, pp. 1411--1416.

\bibitem{mann-kendall_2017}
I.~Umesh, G.~Srinivasan, and M.~Torquato, ``Software aging forecasting using
  time series model,'' \emph{Indonesian Journal of Electrical Engineering and
  Computer Science}, vol.~7, no.~3, pp. 839--845, 2017.

\bibitem{mann_kendall_confidence_95_2013}
F.~Machida, A.~Andrzejak, R.~Matias, and E.~Vicente, ``On the effectiveness of
  mann-kendall test for detection of software aging,'' in \emph{2013 IEEE
  International Symposium on Software Reliability Engineering Workshops
  (ISSREW)}.\hskip 1em plus 0.5em minus 0.4em\relax IEEE, 2013, pp. 269--274.

\bibitem{statistical_1987}
R.~O. Gilbert, \emph{Statistical methods for environmental pollution
  monitoring}.\hskip 1em plus 0.5em minus 0.4em\relax John Wiley \& Sons, 1987.

\bibitem{rejuvcomparative}
J.~Alonso, R.~Matias, E.~Vicente, A.~Maria, and K.~S. Trivedi, ``A comparative
  experimental study of software rejuvenation overhead,'' \emph{Performance
  Evaluation}, vol.~70, no.~3, pp. 231--250, 2013.

\bibitem{openstack_aging_mem_usage_component_melo_2017}
C.~Melo, J.~Araujo, V.~Alves, and P.~R.~M. Maciel, ``Investigation of software
  aging effects on the openstack cloud computing platform.'' \emph{J. Softw.},
  vol.~12, no.~2, pp. 125--137, 2017.

\bibitem{openstack_failures_2019}
D.~Cotroneo, L.~De~Simone, P.~Liguori, R.~Natella, and N.~Bidokhti, ``How bad
  can a bug get? an empirical analysis of software failures in the openstack
  cloud computing platform,'' in \emph{Proceedings of the 2019 27th ACM Joint
  Meeting on European Software Engineering Conference and Symposium on the
  Foundations of Software Engineering}, 2019, pp. 200--211.

\bibitem{command_design_patterns}
E.~Gamma, R.~Helm, R.~Johnson, and J.~Vlissides, ``Design patterns: Abstraction
  and reuse of object-oriented design,'' in \emph{ECOOP’93—Object-Oriented
  Programming: 7th European Conference Kaiserslautern, Germany, July 26--30,
  1993 Proceedings 7}.\hskip 1em plus 0.5em minus 0.4em\relax Springer, 1993.

\bibitem{mann_kendall_web_server_2006}
M.~Grottke, L.~Li, K.~Vaidyanathan, and K.~S. Trivedi, ``Analysis of software
  aging in a web server,'' \emph{IEEE Transactions on reliability}, vol.~55,
  no.~3, pp. 411--420, 2006.

\bibitem{sware_mann_kendall_2018}
M.~Torquato, J.~Araujo, I.~Umesh, and P.~Maciel, ``Sware: a methodology for
  software aging and rejuvenation experiments,'' \emph{Journal of Information
  Systems Engineering and Management}, vol.~3, no.~2, p.~15, 2018.

\bibitem{mann_kendall_aging_cloud_2020}
E.~Andrade, F.~Machida, R.~Pietrantuono, and D.~Cotroneo, ``Software aging in
  image classification systems on cloud and edge,'' in \emph{2020 IEEE
  International Symposium on Software Reliability Engineering Workshops
  (ISSREW)}.\hskip 1em plus 0.5em minus 0.4em\relax IEEE, 2020, pp. 342--348.

\bibitem{aging_mann_kendall_sens_slope}
K.~S. Trivedi, K.~Vaidyanathan, and K.~Goseva-Popstojanova, ``Modeling and
  analysis of software aging and rejuvenation,'' in \emph{Proceedings 33rd
  annual simulation symposium (SS 2000)}.\hskip 1em plus 0.5em minus
  0.4em\relax IEEE, 2000, pp. 270--279.

\bibitem{mann_kendal_fundamentals_1945}
H.~B. Mann, ``Nonparametric tests against trend,'' \emph{Econometrica: Journal
  of the econometric society}, pp. 245--259, 1945.

\bibitem{sen1968estimates}
P.~K. Sen, ``Estimates of the regression coefficient based on kendall's tau,''
  \emph{Journal of the American statistical association}, vol.~63, no. 324, pp.
  1379--1389, 1968.

\bibitem{moving_averages_advantage}
R.~Matias, A.~Andrzejak, F.~Machida, D.~Elias, and K.~Trivedi, ``A systematic
  differential analysis for fast and robust detection of software aging,'' in
  \emph{2014 IEEE 33rd International Symposium on Reliable Distributed
  Systems}.\hskip 1em plus 0.5em minus 0.4em\relax IEEE, 2014, pp. 311--320.

\bibitem{moving_average_example_2017}
I.~Umesh, G.~Srinivasan, and M.~Torquato, ``Software aging forecasting using
  time series model,'' \emph{Indonesian Journal of Electrical Engineering and
  Computer Science}, vol.~7, no.~3, pp. 839--845, 2017.

\bibitem{kendall_book_1948}
M.~G. Kendall, ``Rank correlation methods.'' 1948.

\end{thebibliography}

\section{Appendix}~\label{appendix}
In the following, we give the details of the implemented testbed and the experimental design.
\subsection{OpenStack testbed}

Several tools are available for conducting tests on OpenStack; however, there is no single solution capable of deploying OpenStack, applying accelerated testing load, and collecting the necessary data. Both the Rally and Yardstick projects partially meet the requirements, but neither can deploy OpenStack. While the Rally project previously deployed OpenStack using DevStack, this functionality is no longer supported. Consequently, we opted to use the following tools for our research:

\begin{itemize}
    \item Kolla-Ansible for the deployment and removal of OpenStack;
    \item Rally-OpenStack, also known as Rally, for executing workloads and gathering information on their performance;
    \item Prometheus for collecting data on memory consumption;
    \item Flask framework, which was used to develop web application for user interface and automating scenario execution.
\end{itemize}

Utilizing these components, we developed and implemented a testbed for conducting OpenStack experiments. The computers within our testbed were equipped with the following specifications: Intel(R) Xeon(R) CPU E3-1230 V2 @ 3.30GHz with 8 cores, 16 GB of DDR3 1333 Mhz RAM, and 16 GB of swap space, all running on Ubuntu 18.04.5 LTS.

Our experimental environment was configured with the following module releases: OpenStack Victoria, kolla-ansible 11.1.0, and ansible 2.9.0. A total of five machines were utilized in this experiment: one served as the experiment orchestration center, with the Flask web application deployed on it, while the remaining four machines functioned as host machines for our cloud deployments. All five machines were interconnected within the same network, with the orchestration machine possessing root access to the other four. Prior to its startup, the web application received configuration information about the working environment in the form of a configuration file, which included a comprehensive list of all orchestrated computers considered as OpenStack node candidates, complete with their respective IP and domain names.
In the following we give details on the inner of the four components. 

\subsubsection{Flask Application}
The Flask application consists of the following modules:

\begin{itemize}
\item The web application module, which is responsible for providing users with a web interface, processing user input, and storing it as requests.
\item The request manager module, which schedules user requests and initiates their execution in separate threads, each dedicated to distinct OpenStack deployments.
\item The OpenStack management module, is responsible for executing requests associated with managing the life cycle of OpenStack.
\item The experiment module, is responsible for requests linked to the generation and execution of experiments on OpenStack, as well as the collection of essential data.
\end{itemize}

\subsubsection{User Interface}
When a user issues a specific action to be performed on the OpenStack cloud via the web application interface, this module parses the input data and generates a request for the request manager to schedule, effectively acting as an intermediary layer between the user and the request manager submodule. All requests, that are scheduled for execution, can also be cancelled via the user interface. Prior to scheduling, the request data are stored in an application database, enabling their reuse for subsequent requests with either identical or modified parameters. This configuration flexibility allowed us to set the workload parameters once and subsequently replicate the request for execution, even across various deployments.

Web application submodule is able to process and pass to the request manager the request types:
\begin {itemize}
    \item Prepare OpenStack deployment files;
    \item Deploy OpenStack;
    \item Destroy OpenStack;
    \item Delete OpenStack configuration;
    \item Run Rally workload;
    \item Clean OpenStack;
    \item Download OpenStack metrics;
\end {itemize}

\begin{figure}[!t]
\centering
\includegraphics[width=0.48\textwidth]{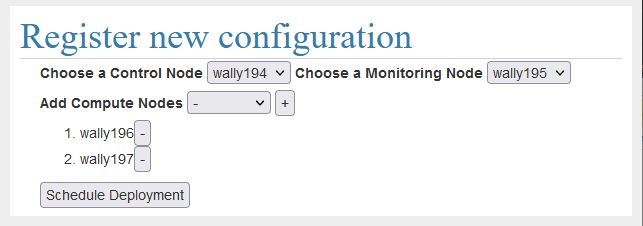}
\caption{OpenStack installation request page.}
\label{figure/chapter_5/testbed_deploy_screenshot}
\end{figure}

The web application submodule enables users to define an OpenStack configuration consisting of nodes, that are not occupied by other OpenStack clouds. Figure~\ref{figure/chapter_5/testbed_deploy_screenshot} shows an example of what the multi-node OpenStack installation request looks like in a web application.

A single OpenStack configuration requires the definition of a single control node, a single monitoring node, and at least one compute node, with the possibility of assigning to a node a combination of roles. 

For the all-in-one configuration, we assigned all three nodes to a single machine. For the multi-node configuration, we used a configuration with one control node, one monitoring node, and two compute nodes. The separation of the monitoring node from the control one allows us to distinguish ageing in the central OpenStack control node from that in the optional monitoring node, contributing to a more realistic setup, as real systems often employ separate monitoring systems. The use of two distinct compute nodes enabled the observation of ageing in the control node, which does not itself perform compute node functions. Additionally, the ability to distribute the workload between these two compute nodes brought to the cloud its main advantage of being able to distribute the load and thus, contrasted the all-in-one and multi-node setups explicit, particularly at high concurrency levels.

\subsubsection{Request Manager}
\begin{figure}[!t]
\centering
\includegraphics[width=0.48\textwidth]{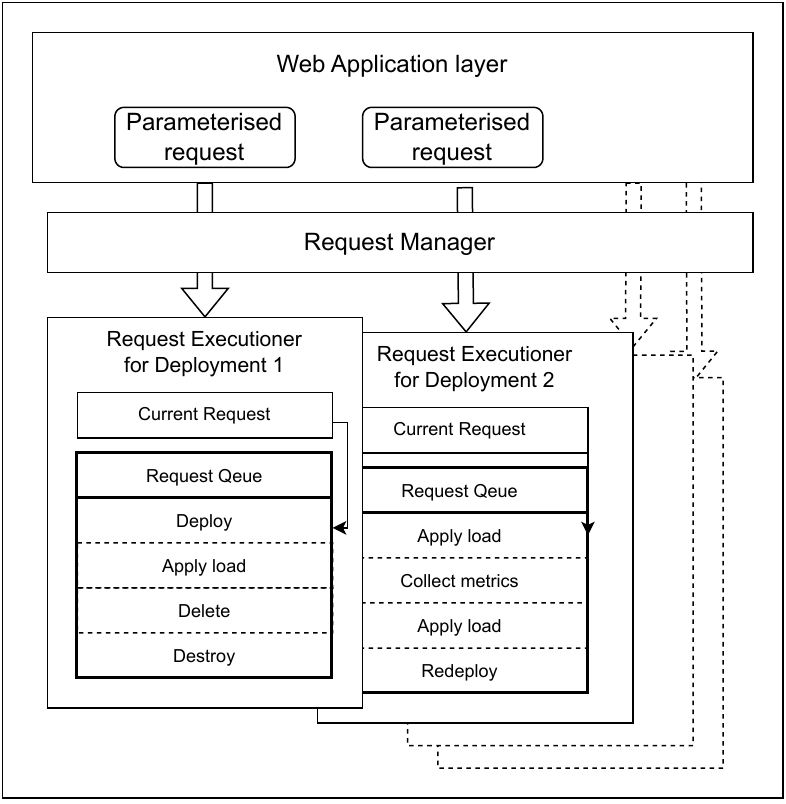}
\caption{Request manager workflow scheme.}
\label{figure/chapter_5/request_manager}
\end{figure}
When request manager receives a request, it first stores it into an application database and then schedules it for execution. For each OpenStack deployment, a distinct request execution thread is created, which processes scheduled requests one by one, thus ensuring that requests for different OpenStack deployments can be executed at the same time without interference with each other (Figure~\ref{figure/chapter_5/request_manager}).

\subsubsection{OpenStack's Management Module}
OpenStack's management module is responsible for the execution of requests, associated with the OpenStack lifecycle, such as:

\begin{itemize}
    \item \texttt{Prepare OpenStack} deployment files;
    \item \texttt{Deploy OpenStack};
    \item \texttt{Destroy OpenStack};
    \item \texttt{Delete OpenStack}.
\end{itemize}

To execute these operations, the package \texttt{kolla-ansible} is used, which is called using Python's \texttt{subprocess} library. During the \texttt{Prepare OpenStack} request, configuration files for the future OpenStack deployment are generated. These files define which OpenStack modules will be installed, as well as the assignment of OpenStack node roles, among other configurations. The \texttt{Deploy OpenStack} request ensures that the necessary packages are installed on the target machines, generates additional deployment files, and deploys OpenStack on these machines in the form of Docker containers. The \texttt{Destroy OpenStack} request is responsible for tearing down the OpenStack deployment by stopping and deleting the Docker containers. Lastly, the \texttt{Delete OpenStack} operation erases all configuration files and data generated during the runtime of OpenStack.

\subsubsection{Experiment module}
The Experiment module oversees the process of executing experiments and collecting data. It receives experiment request data from the request manager, overseeing two key functions utilized in this study: the execution of OpenStack workloads and the collection of metrics.

\subsubsection{Workload Execution} \label{workload}
The OpenStack workload request operates as a decorator around the rally-OpenStack task execution. Its primary parameter is the workload task body, which defines the specific rally workload to be executed, including workload parameters.

For this study, a custom workload task was implemented in the form of a rally-OpenStack plugin. This workload task represents a sequence of OpenStack atomic actions that are executed in the following order: create user, create role, add role, create security group, create flavor, create image, create network, create subnet, create port, create router, boot server, create volume, attach volume, rebuild server, pause server, unpause server, detach volume, delete volume, delete server, delete router, delete port, delete subnet, delete network, delete image, delete flavor, delete security group, revoke role, delete role, delete user. During this workload different OpenStack modules are subjected to load, such as nova, neutron, keystone, glance and cinder. Each of the workload actions is executed by one of the modules within OpenStack, and some of these actions depend on others. The categorization based on the modules and their interdependencies can be observed in Table~\ref{table:openstack_actions}.

\begin{table}
\begin{center}
\resizebox{0.47\textwidth}{!}{%
\begin{tabular}{ |c|c|c| } 
 \hline
 OpenStack module & OpenStack action & Action dependency \\ 
\hline
\hline
\multirow{3}{4em}{Keystone} & create/delete user &  \\\cline{2-3}
& create/delete role  &  \\\cline{2-3}
& add/revoke role & keystone.role,nova.user \\ 
\hline
\hline
\multirow{6}{4em}{Neutron} & create/delete security group &  \\\cline{2-3}
& create/delete network  &  \\\cline{2-3}
& create/delete subnet  & neutron.network  \\\cline{2-3}
& create/delete port  & neutron.network  \\\cline{2-3}
& create/delete port  & \\\cline{2-3}
\hline
\hline
Glance & create/delete image&\\
\hline
\hline
Cinder & create/delete volume&\\
\hline
\hline
\multirow{6}{4em}{Nova} & create/delete flavor &  \\\cline{2-3}
& create server  & nova.flavor, glance.image,\\
& delete server & neutron.network \\\cline{2-3}
& attach/detach volume  & nova.server, cinder.volume  \\\cline{2-3}
& pause/resume server  & nova.server  \\\cline{2-3}
& rebuild server & nova.server, glance.image \\\cline{2-3}
 \hline
\end{tabular}
}
\end{center}
\caption{List of OpenStack modules, actions and their dependencies, that were used in workload definition}
\label{table:openstack_actions}
\end{table}

During the implementation of our workload structure, we used the command design pattern to facilitate the integration of the clean-up procedure \cite[p.~233]{command_design_patterns}. Taking into account the command pattern principle, a request to execute a workload step was encapsulated as an object, enabling dynamic scheduling and interruption capabilities. This design allowed us to establish a one-to-one relationship between each creation step and its corresponding deletion counterpart, encapsulating both within the command object. Additionally, the rebuild and pause server actions were included as creation steps within the command objects, even in the absence of direct counterparts.

Within the workload execution phase, each creation action was executed based on its position in the workload definition, and its associated command object was added to a Last In, First Out queue. Upon completion of all creation actions, the command objects were retrieved from the queue, and the corresponding deletion actions were triggered for each. This process allowed the workload to delete all entities created during its initial phase. In the event of an exception occurring during any workload step, the execution was directed to the deletion stage, resulting in the invocation of the deletion action for all decorators for which the creation action had already been successfully executed. This approach facilitates easy extension and modification of the workload structure while ensuring that the clean-up mechanism automatically adapts to any changes.

\subsubsection{Data Collection}
Rally supports the generation of rally reports during the execution of requested workloads, a feature that we integrated into our web application. This capability allowed us to monitor the progress of scenario execution in real time and reset scenario execution if necessary. Given the extended duration of our scenario, this real-time monitoring feature proved especially advantageous, saving significant time in instances where the execution deviated from the intended plan at the start of execution. Upon the completion of a scenario, two rally reports are generated - one for the ageing phase and another for the post-rejuvenation phase. 

These rally reports offer comprehensive insights into the execution of each workload, including the specific start and end times for the entire workload and each individual step within it. In the case of any failures during a workload, the reports also include an error log, providing valuable information about the specific stage within our customized workload where the error occurred.

As stated earlier, we segment the ageing phase into 24 one-hour intervals and allocate workloads to the respective intervals based on their starting timestamps. Subsequently, we calculate the arithmetic mean to determine the average workload duration for each hour of the ageing phase and the rejuvenation phase. These averaged values are utilized in the computation of the data outlined in the methodology section.

After gathering workload execution data, we proceed with the collection of the memory usage metrics. In this regard, we analyze the data obtained from the Prometheus module. Kolla-ansible module deploys data endpoints on every node in OpenStack, each of which provides access to up-to-date node state data via http access. The Prometheus module itself is deployed on the monitoring node, and configured to gather data from endpoints and store them at 30-second interval. Upon initiating the request to collect metrics from Prometheus, a collection of scripts is activated, enabling the retrieval of the stored metrics through the Prometheus HTTP API. The metrics provided by Prometheus are represented as a time series, which is essentially a sequence of observations arranged chronologically. After collection, metric values are allocated by their timestamps either to one of the 24 one-hour intervals or to the rejuvenation phase. Then for each metric, an arithmetic mean is calculated and used in evaluation.

\subsection{Software Ageing Evaluation}~\label{stats}
The predominant characteristics of the software ageing effect on ageing indicators include its gradual escalation, negative impact on a system, persistent presence over time, and a notable reduction upon the implementation of rejuvenation actions.

To assess whether software ageing indicators are susceptible to these effects, we conduct a comprehensive software ageing evaluation. The Mann-Kendall test is employed to detect the presence of a trend and its direction, providing insights into whether the changes in the indicator are negative and of a long-term nature. Utilizing Sen's Slope allows us to estimate the growth rate of the trend, facilitating comparisons between the same indicators in different contexts and enhancing our understanding of how the context influences indicator behaviour. Additionally, calculations based on the fundamental definition of ageing effect aid in estimating the cumulative impact of ageing and assessing the effectiveness of the rejuvenation procedure. This comprehensive approach enables us to verify or question the findings derived from the other two evaluation methods.

A combination of the Mann-Kendall test and Sen's slope is often used as a statistical method to detect and evaluate trends of software ageing in ageing indicators \cite{mann_kendall_web_server_2006, sware_mann_kendall_2018, mann_kendall_aging_cloud_2020, mann-kendall_2017, aging_response_hosted_app_2021, aging_mann_kendall_sens_slope}.  The Mann-Kendall test checks the null hypothesis, \texttt{H0}, which states that there is no trend in the data during the period, against the alternative hypothesis, \texttt{H1}, which indicates an upward or a downward monotonic trend in the data \cite{mann_kendal_fundamentals_1945}. If the Mann-Kendall test detects the presence of a trend, the slope of the monotonic trend is calculated using Sen's slope \cite{sen1968estimates}. To run the Mann-Kendall test, we first have to gather historical data on ageing indicators of the OpenStack system over some time. This data creates the time series data. We employ hourly data by calculating moving averages for each hour. It provides a practical balance between data reduction and trend visibility, by smoothing short-term fluctuations and highlighting longer-term trends, enhancing the overall interpretability \cite{moving_averages_advantage, moving_average_example_2017}. After gathering data, the Mann-Kendall test statistic must be calculated, which quantifies the strength and direction of the trend in the data. The test statistic takes into account the differences between pairs of data points and assesses their distribution. In our case, the test statistic is calculated as follows \cite[p.~208-212]{statistical_1987}

\[
S = \sum_{i=1}^{n-1} \sum_{j=i+1}^{n} \text{sgn}(x_j - x_i)
\]
Where \texttt{S} is the number of positive differences minus the number of negative differences,  \texttt{n} is the number of data points, $x_i$ and $x_j$ are the value of the time series at position \texttt{i} and \texttt{j} respectfully, and \texttt{sgn(a)} is the sign function, which returns  -1  if $a < 0$, 0  if $a = 0$, and 1 if $a > 0$. If S is positive, it means that later observations tend to be larger than earlier ones. If S is negative, it means that later observations tend to be smaller than earlier ones. If n is greater than 10 then we calculate the variance of S, which is then used to calculate another test statistic Z \cite[p.~55]{kendall_book_1948}:

%
%
              
where:

\begin{align*}
& - Z \text{ is the standard normal (Z) score;} \\
& - S \text{ is the Mann-Kendall test statistic;} \\
& - \text{var}(S) \text{ is the variance of the Mann-Kendall test statistic.} \\
\end{align*}
The variance of the Mann-Kendall test statistic:
\[
\text{var}(S) = \frac{n(n-1)(2n+5) - \sum_{i=1}^{k} t_i(t_i-1)(2t_i+5)}{18}
\]

where:
\begin{align*}
& - \text{var}(S) \text{ is the variance of the Mann-Kendall test statistic;} \\
& - n \text{ is the number of data points;} \\
& - k \text{ is the number of tied groups in the data;} \\
& - t_i \text{ is the number of data points in the } i\text{-th tied group.} \\
\end{align*}

This test helps determine whether the observed trend is statistically significant or if it could have occurred by chance. The significance level is compared to the quantiles of the standard normal distribution to check if we accept the hypothesis that there is no trend. If the calculated significance level is below a chosen threshold (e.g., 0.05), then we can conclude that there is a significant trend in the data. If the significance level is above the threshold, we cannot confidently say that there is a trend. Applying the Mann-Kendall Test to the performance and memory usage data of OpenStack allows us to assess whether there is a consistent upward or downward trend over time, potentially indicating software ageing. We can also compare which criteria of significance the results of calculations met, but to do it in a more efficient way we are going to use Sen's slope.

Sen's slope estimate is a robust statistical method used to determine the rate of change or slope in a dataset, particularly in the presence of outliers or non-normally distributed data \cite{sen1968estimates}.
For the set of pairs $(i, x_i)$, where $x_i$ is a time series, Sen's slope estimate is computed in the following way:
\[
\texttt{Sen's slope} = Median\{\frac{x_j-x_i}{j-i}:i<j\}
\]

At the significance level of the test $\alpha = 0.05$, the hypothesis of absence of trend is rejected, if $|Z| > 1.96$, the trend is upward if $Z > 1.96$ and the trend is downward if $Z < 1.96$. A positive Sen's slope denotes a positive directional trend, while a negative Sen's slope indicates a negative directional trend, with the magnitude of the slope signifying the rate of change of the trend, whether ascending or descending.

\end{document}